\documentclass[prb,aps,twocolumn,preprintnumbers,amsmath,amssymb,superscriptaddress,altaffilletter]{revtex4}

\usepackage[utf8]{inputenc}
\usepackage{array}
\usepackage{amsmath}
\usepackage{multirow}
\usepackage{graphicx}
\usepackage{dcolumn}

\begin{document}

\preprint{PREPRINT}
\title{Quantum Cascade Laser Frequency Combs}
\author{J\'er\^ome Faist}
\author{Gustavo Villares}
\author{Giacomo Scalari}
\author{Markus R\"osch}
\author{Christopher Bonzon}
\affiliation{Institute for Quantum Electronics, ETH Z\"urich, Switzerland, E-mail: jfaist@ethz.ch}
\author{Andreas Hugi}
\affiliation{IRsweep GmbH, Z\"urich, Switzerland, E-mail: andreas.hugi@irsweep.com}
\author{Mattias Beck}
\affiliation{Institute for Quantum Electronics, ETH Z\"urich, Switzerland, E-mail: jfaist@ethz.ch}

\date{\today}
%
%
%
%
%
%


  \begin{abstract}
{It was recently demonstrated that broadband quantum cascade lasers can operate as frequency combs. As such, they operate under direct electrical pumping at both mid-infrared and THz frequencies, making them very attractive for dual-comb spectroscopy. Performance levels are continuously improving, with average powers over 100\,mW and frequency coverage of 100\,cm$^{-1}$ in the mid-infrared. In the THz range, 10\,mW of average power and 600\,GHz of frequency coverage are reported. As a result of the very short upper state lifetime of the gain medium, the mode proliferation in these sources  arises from four wave mixing rather than saturable absorption. As a result, their optical output is characterized by the tendency of small intensity modulation of the output power, and the relative phases of the modes to be similar to the ones of a frequency modulated laser. Recent results include the proof of comb operation down to a metrological level, the observation of a Schawlow-Townes broadened linewidth, as well as the first dual-comb spectroscopy measurements. The capability of the structure to integrate monolithically non-linear optical element as well as to operate as a detector show great promise for future chip integration of dual-comb systems.}
\end{abstract}
  \keywords{spectroscopy, mid-infrared, dual-comb, multiheterodyne}

 
  \startpage{1}


\maketitle

\section{Introduction}

\begin{figure}[!htb]
	\includegraphics[width=0.45\textwidth]{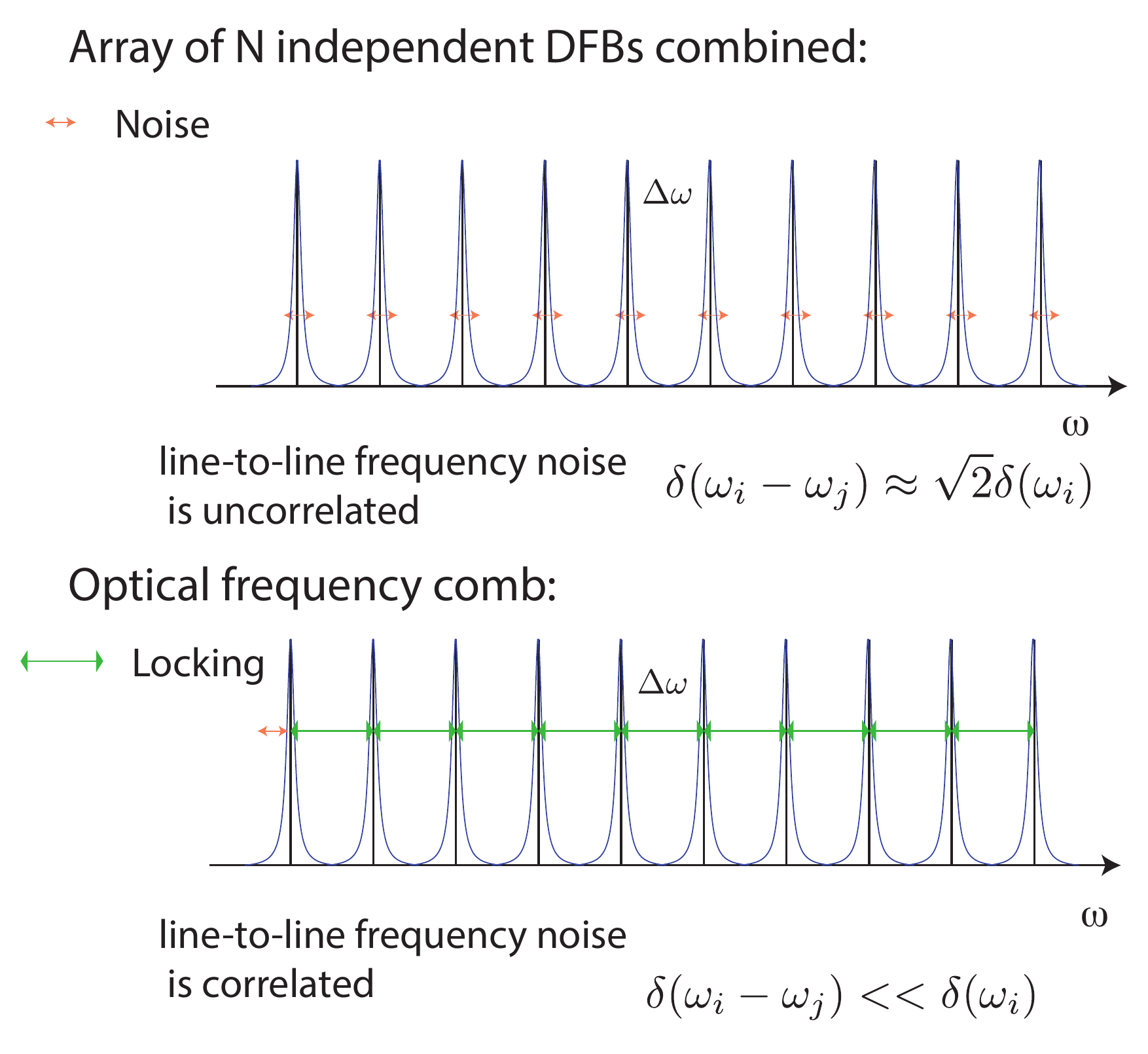}
		\caption{Schematic difference between an array of single mode lasers (top) and a frequency comb (bottom). }
		\label{fig:whatisacomb}
	\end{figure}

An optical frequency comb~\cite{Udem:2002p2021} is an optical source with a spectrum constituted by a set of modes which are 
perfectly equally spaced and have a well-defined phase relationship between each other. Frequency combs can be obtained by exploiting the natural phase locking mechanism arising when lasers operate in mode-locking regime producing ultrafast pulses. As a result, the ensemble of comb frequency lines at frequencies $f_n$, given by 
\begin{equation}
f_n = f_{\textrm{ceo}} + n f_{\textrm{rep}}
\label{eq:fundcomb}
\end{equation}
 are spaced by the repetition rate of the laser $f_{\textrm{rep}}$ which can be made extremely stable and locked to an external microwave source. In addition, the ensemble of frequency lines can be shifted by the so-called carrier-envelope offset frequency $f_{\textrm{ceo}}$~\cite{Udem:2002p2021}. As illustrated in Fig.\,\ref{fig:whatisacomb}, the  difference between an array of single frequency lasers and an optical comb is the correlation in the noise of each individual line, immediately apparent when considering noise terms added to $f_{\textrm{ceo}}$ and to $f_{\textrm{rep}}$. Whereas the heterodyne beat between two independent single mode optical sources with similar linewidth $\delta f_n$ will yield a signal with a linewidth $\delta \nu_{RF} = \sqrt{2} \delta f_n$, the same experiment performed on a comb will yield a value that may be much below the one of the individual lines because of the correlation between the noise of the lines. 
This very peculiar relationship between the modes enables the concept of {\em self-referencing}. In a mode-locked laser broadened to more than an octave~\cite{Udem:2002p2021}, by beating the second harmonic of a line in the red portion with a line in the blue portion of the spectrum, the offset frequency $f_{\textrm{ceo}}$ of the comb can be directly retrieved and stabilized~\cite{Telle:1999cq}. As a result, the absolute optical frequency of each comb line is rigidly linkeCundiff:2003umd to the microwave reference frequency $f_{\textrm{rep}}$, allowing optical frequency combs to act as rulers in the frequency domain. By enabling extremely accurate frequency comparisons using a direct link between the microwave and optical spectral ranges, frequency combs have opened new avenues in a number of fields, including fundamental time metrology \cite{Cundiff:2003um,Udem:2002p2021}, spectroscopy as well as frequency synthesis. In addition, they also had a tremendous impact on many other fields such as astronomy, molecular sensing, range finding, optical sampling, and low phase noise microwave generation~\cite{Diddams:2010p2046}. Their fundamental significance and impact on science was reflected by the attribution of the Nobel Prize in Physics in 2005 to Theodor W. H{\"a}nsch and John L. Hall~\cite{_nobelprize.org._2013}. 

Besides mode-locked lasers, optical frequency combs have recently also been generated using high-Q microcavity resonators pumped by narrow linewidth continuous wave lasers~\cite{Delhaye:2007p1571,Kippenberg:2011p2042}. In this case the Kerr non-linearity is  responsible for establishing the stable phase relationship between the laser modes. In contrast to mode-locked lasers, Kerr combs can exhibit complex phase relations between modes that do not correspond to the emission of single pulses while remaining highly coherent\cite{DelHaye:2014de}. Operation of a Kerr comb in a pulsed regime with controlled formation of temporal solitons was recently demonstrated~\cite{Herr:PRLsolitons:2014}.  

\begin{figure}[!htb]
		\includegraphics[width=0.45\textwidth]{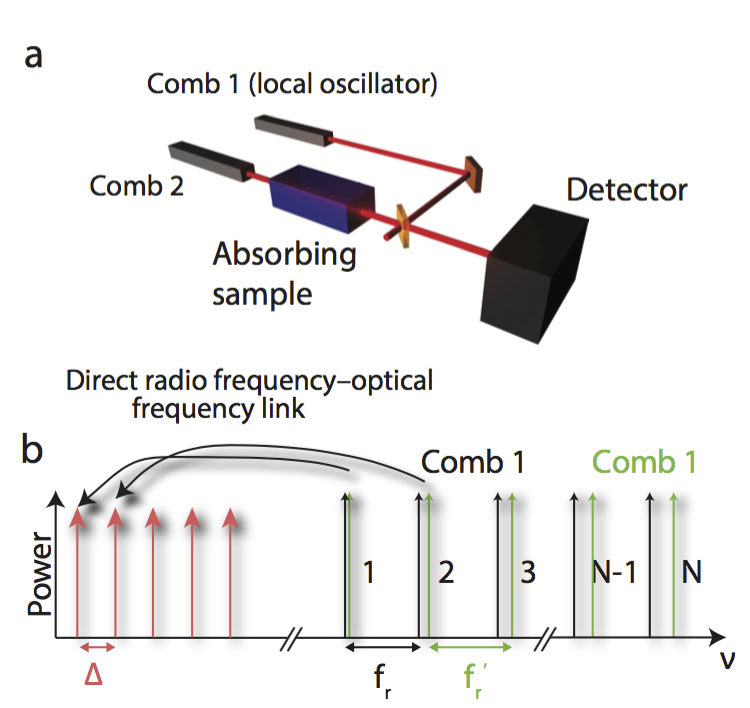}
		\caption{Principle of dual-comb spectroscopy.
			\textbf{a} Schematic diagram of a dual-comb spectroscopy setup. 
			\textbf{b} Schematic diagram in the frequency domain.}
		\label{fig:dualcomb}
	\end{figure}

An extremely appealing application of optical frequency combs is the so-called dual-comb spectroscopy, where multi-heterodyne detection is performed allowing Fourier transform spectroscopy with potentially high resolution, high sensitivity and no moving parts~\cite{Keilmann:2004p2040,Coddington:2008p99}. As shown schematically in Fig.\,\ref{fig:dualcomb}, two combs with slightly different repetition rates are used  in a local oscillator-source configuration. Indeed, spectroscopy by means of optical frequency combs surpassing the precision and speed of traditional Fourier spectrometers by several orders of magnitude was recently demonstrated~\cite{baumann_spectroscopy_2011,rieker_open-path_2013,leindecker_octave-spanning_2012}. 
%
The use of dual-comb spectroscopy is especially interesting in the mid-infrared portion of the spectrum, the so-called ``molecular fingerprint region'' where most fundamental roto-vibrational absorption bands of light molecules can be found. Applications of mid-infrared spectroscopy are in the areas of  environmental sensing, including isotopologues, medical, pharmaceutical, and toxicological measurements as well as homeland security applications for molecules that are related to explosives. The THz region is also of high importance for non-invasive imaging, astronomy, security and medical applications~\cite{Tonouchi:2007p1411,Spagnolo:2015ho}. For these reasons, there is a very strong demand to also create optical frequency combs centered in the mid-infrared and THz regions of the spectrum~\cite{Schliesser:2012dn}. A possible approach is to downconvert the near-infrared emission by means of non-linear optics, still maintaining their short-pulse nature \cite{Adler:OPEX:2009}. Direct mid-infrared emission, down to 6.2$\mu$m wavelength, has been also achieved using direct pumping of an OPO using a Tm-doped fiber laser~\cite{Leindecker:2012ed}. Also very interesting are the results achieved using high-Q microresonators based on Si waveguides \cite{Griffith:2015gd,Kuyken:2015bw} or microcrystalline resonators~\cite{Wang:2013db}. Recently, such a comb was generated in a microresonator using optical pumping from a quantum cascade laser (QCL) at $\lambda \approx 4.5 \mu$m \cite{Lecaplain:2015wf,Savchenkov:2015ha}. Nevertheless, one common drawback of these optical sources is that they consist of different optical elements that must be assembled together. 

This paper reviews an emerging new optical frequency comb technology based on QCLs. Section 2 reviews the basic physics, as well as the main operation characteristics of QCL combs. In section 3, the techniques developed to characterize the comb operation are described. The topic of dispersion compensation is addressed in section 4. The application in spectroscopy of dual-combs systems based on QCL combs is discussed in section 5. Conclusions and outlook are reported in section 6. 

\section{QCL combs}
\paragraph*{QCL broadband technology}
QCLs~\cite{Faist:2013td} are semiconductor injection lasers emitting throughout the mid-infrared (3-24\,$\mu$m) and THz (50-250\,$\mu$m)~\cite{Kohler:2002p64,Scalari:2009p746} regions of the electromagnetic spectrum. 
First demonstrated 
in 1994~\cite{Faist:1994p1420} at $\lambda \approx 4.3 \mu$m, they have undergone a tremendous development. Their capability to operate in a very wide frequency range makes them very convenient devices for optical sensing applications. In particular, single frequency emitter devices have demonstrated both very large continuous optical output power up to 2.4\,W~\cite{Lu:2011ir}, high temperature operation in continuous wave~\cite{Wittmann:2009p1422} and low electrical dissipation below a Watt~\cite{Bismuto:2015gg}. The physics of quantum cascade laser is, in addition, very beneficial for their operation as broadband gain medium. First, intersubband transitions are transparent on either side of their transition energy. Interband transitions, in contrast, are transparent only on the low-frequency side of their gain spectrum and highly absorbing on the high-frequency side. Second, the cascading principle almost comes naturally because of the unipolar nature of the laser. These two features enable the cascading of dissimilar active region designs emitting at different wavelengths to create a broadband emitter. This concept, first demonstrated at cryogenic temperature~\cite{Gmachl:2002p63} was further developed for applications with high performance, inherently broad gain spectra designs where a single upper state exhibits allows transitions to many lower states~\cite{Hugi:2009p1437}. This technology was the base to the fabrication of external cavity quantum cascade lasers with very large tunability~\cite{Hugi:2010ca}. 

\paragraph*{Mode-locking of QCLs}
It was suggested very early that broadband QCL could be mode-locked to provide ultrashort mid-infrared pulses~\cite{Gmachl:2002p63}. 
	\begin{figure*}[t]
		\centering
 		\includegraphics[width=15cm]{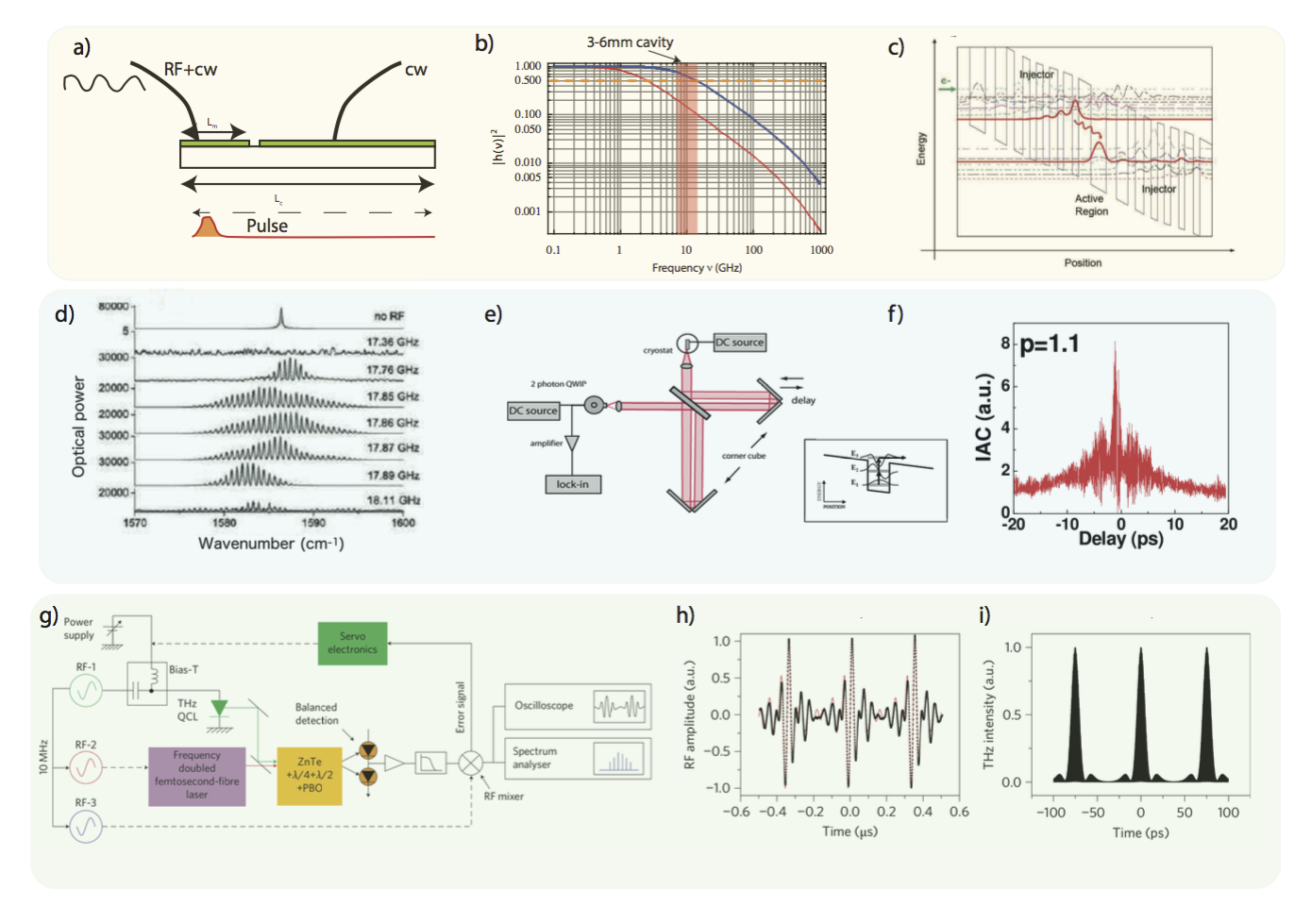}
		\caption{Mode-locking QCLs. 
			\textbf{a} Schematic diagram of an actively mode-locked QCL: the modulated section opens a window in time, allowing a single pulse to propagate in the cavity.
			\textbf{b} The computed transfer function of a QCL, showing the high modulation capability at frequencies corresponding to the round-trip frequency of a 3-6\,mm cavity laser. Results published in~\cite{Faist:2013td}.
			\textbf{c} Active region based on a photon-assisted tunneling transition, with very long upper state lifetime. Results published in~\cite{Wang:2009p1554}.  
			\textbf{d} Spectrum of the device presented in \textbf{c} under modulation.
			\textbf{e} Interferometric autocorrelation technique: the optical pulses are characterized by Fourier Transform Interferometer followed by a two-photon quantum well photodetector.
			\textbf{f} The resulting interferogram, measured 10\% above threshold, shows the ratio of 8:1 between the center and the wings, expected for single pulses. Results published in~\cite{Wang:2009p1554}.
			\textbf{g} Electro-optic sampling, using a femto laser comb of the output of an actively mode-locked THz QCL. Results published in~\cite{barbieri:2011p1986}.
			\textbf{h} THz reconstructed electric field trace. Results published in~\cite{Barbieri:2010p1763}.
			\textbf{i} Intensity of the pulses. Results published in~\cite{Barbieri:2010p1763}.
			}
		\label{fig:modelocked}
	\end{figure*}
In a mode-locked laser~\cite{Haus:2000p656}, the fixed phase relationship between the optical modes -- needed for the formation of an optical comb spectrum as described by Eq.~\ref{eq:fundcomb} -- is obtained by having a single pulse propagating in the laser cavity. This mode of operation of the laser is generally obtained either by a subtile compensation between a negative dispersion in the cavity and the Kerr effect, giving rise to the propagation of a temporal soliton, or by a saturable absorber opening a time-window of low loss for a pulsed output. The short pulses that will be produced at the roundtrip frequency of the cavity  will have, in well designed laser, an average power commensurate to the one of a continuous wave laser. This is possible because the energy of the two-level system, stored in the inverted medium, is abruptly released in the optical pulse. In contrast, QCLs, being based on intersubband transitions in quantum wells, exhibit very short upper state lifetime with $\tau_2 \approx$ 1\,ps~\cite{TATHAM:1989p400}. In fact, in high performance devices operating at room temperature, this time is in the sub-picosecond range ($\tau_2 \approx 0.6$\,ps). This time is much shorter than the typical cavity round trip time $\tau_{rt}$ (64\,ps for a 3\,mm long device), such that the product $\omega \tau_2 << 1$, where $\omega = 2 \pi f_{\text{rep}} = 2 \pi / \tau_{rt}$ is the angular frequency corresponding to the longitudinal mode spacing. As a result, passive mode-locking in the sense described above is not possible. 

The very short upper state lifetime is responsible for the very fast dynamics of the transport in QCLs. For this reason, QCLs can be modulated at frequencies above 20 GHz, equal to the round-trip frequency of 3 millimeter long cavities, as shown in Fig.\,\ref{fig:modelocked}\textbf{b}. Active mode-locking is then feasible by modulating one short section of the QCL active region at its round trip frequency (Fig.\,\ref{fig:modelocked}\textbf{a}). However, to mitigate the gain saturation in the active region, it is necessary to work with upper state lifetimes in the tens of picoseconds, such that the condition $\omega \tau_2 \approx 1$ is fulfilled. Based on this principle, active mode-locking has been achieved in the mid-infrared with active regions based on photon-assisted tunnelling transitions (as shown in Fig.\,\ref{fig:modelocked}\textbf{c})~\cite{Wang:2009p1554} where very long intersubband lifetimes in the order of 20\,ps can be achieved~\cite{Faist:1997p493}. At a temperature of 77K, pulse length of 3\,ps with a bandwidth of 15\,cm$^{-1}$ were achieved~\cite{Wang:2009p1554} (see Fig.\,\ref{fig:modelocked}\textbf{c}, \textbf{d} and \textbf{e}). Nevertheless, the gain recovery time of 50\,ps still restricted the operation of the device to a driving current range not much higher than 10\% above threshold, as predicted by numerical simulations~\cite{Gkortsas:2010tu}. 

In the THz, relatively long upper state lifetimes are achieved at cryogenic temperatures with state-of-the art active regions based on bound-to-continuum transitions. Active mode-locking by RF injection was achieved at 3\,THz with pulses lengths of about 10\,ps and a bandwidth of 100\,GHz~\cite{Barbieri:2010p1763,barbieri:2011p1986,Ravaro:ij}, as shown in Fig.\,\ref{fig:modelocked}\textbf{i}. One very interesting aspect of these experiments was the possibility of locking the THz QCL to a near-infrared fiber comb and use the latter to perform electro-optic sampling of the THz pulse (see Fig.\,\ref{fig:modelocked}\textbf{g} and \textbf{h}), as reviewed in~\cite{Sirtori:2013ky}. Finally, mode-locking was also achieved in the THz by a combination of injection seeding and RF gain switching~\cite{Oustinov:2010iu,Freeman:2012hd}. In the latter case, in contrast to the conventional mode-locking by RF injection, the phase of the THz pulse is fixed by seeding the THz QCL laser with the output of a gated photoconductive antenna. 

 However, compared to passive mode-locking in which the pulse length can routinely reach the inverse of the gain bandwidth $\Omega_g$, fundamental active mode-locking leads to much longer pulses since the pulse length $\tau$ is now a geometrical average of the gain bandwidth and the round-trip frequency $\omega$~\cite{Haus:2000p656}. In addition, as mentioned above, the output power is limited by the necessity to remain close to laser threshold.   
%
%
%
All in all, the reduced resulting mode-locking bandwidth and the low optical power therefore limits the usefulness of such devices for broadband spectroscopy. 

Further work in mode-locking QCLs include the use of a ring external cavity. This arrangement simplifies the mode selection because it minimizes the spatial hole burning effect, and it also allows the use of modulation frequencies low enough to enable complete switching of the laser on and off. This geometry has been analysed both theoretically~\cite{Wang:2015gj} and experimentally~\cite{Malara:2013jr,Wojcik:2013fv}. 

Finally, an intriguing proposal is the use of so-called self-induced transparency mode-locking, where an absorber section is precisely tailored to undergo a half-Rabi oscillation under the optical pulse, stabilizing the optical output~\cite{Menyuk:2009p1555}. Although the concept is very general, it was argued that the QCL, due to its short upper state lifetime, is a specially well suited system to implement such mode-locking scheme. The mode-locking was predicted to remain stable even under small amount of dispersion and saturable nonlinearity~\cite{Talukder:2010cy}. Nevertheless, the active region design requires a very fine tuning of the respective oscillator strengths of the gain and absorber sections, and so far no experimental evidence of such mode-locking has been reported. 

\begin{figure}[!htb]
		\includegraphics[width=0.45\textwidth]{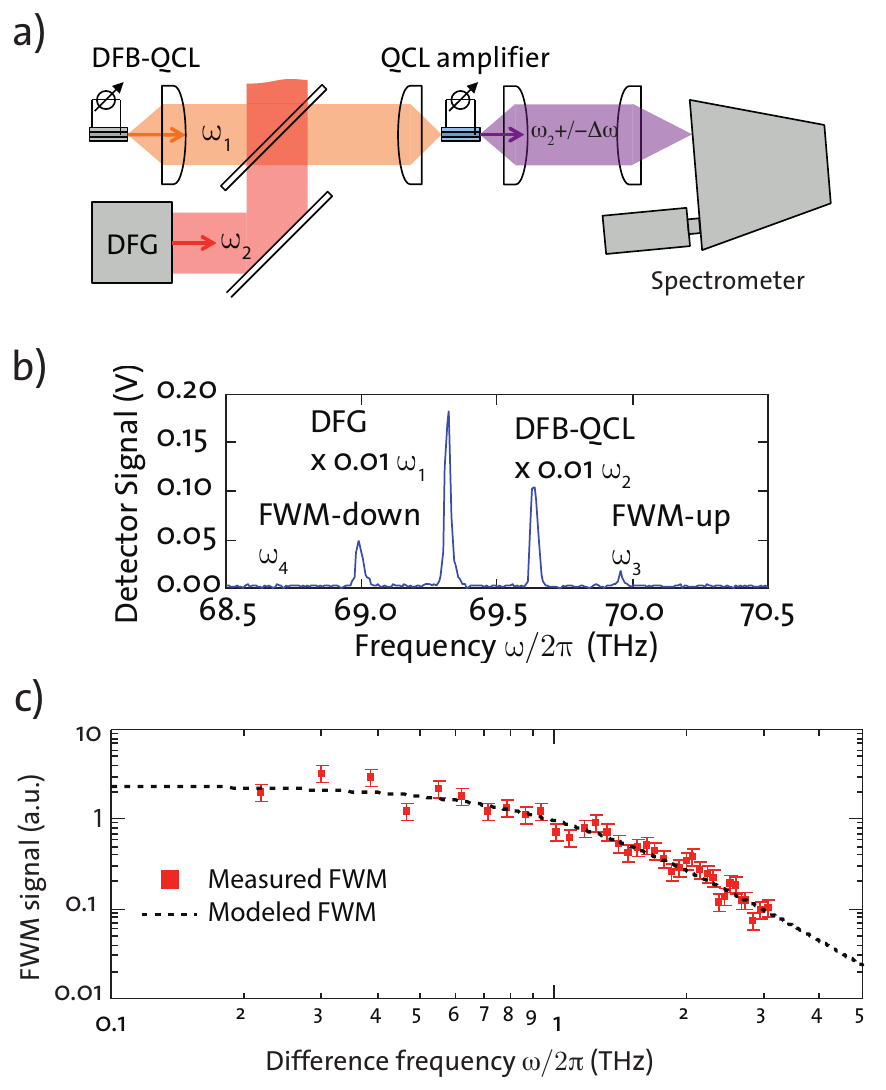}
		\caption{Four wave mixing in a QCL active region. 
			\textbf{a} Schematic diagram of the experimental setup. 
			\textbf{b} Measured spectrum for a source separation $\delta \omega = \omega_2 - \omega_1$ of 314 GHz. 
			\textbf{c} response of the FWM as a function of detuning between the two sources (points), compared with the predictions of a two-level model (dashed lines). Results published in~\cite{Friedli:2013id}.}
		\label{fig:fwm}
	\end{figure}

\paragraph*{Four wave mixing in QCLs}
The active region of the QCL exhibits optical non-linearities, as any two-level system. In particular, the very short upper state lifetime, that strongly hinders fundamental mode-locking, is responsible for a very broadband four wave mixing process. Four wave mixing due to intersubband transition was first measured in a doped multiquantum well~\cite{Walrod:1999p1880}. In a QCL active region under operation, measurements of the four wave mixing (FWM) product was performed by injecting two single frequency sources and analysing the output spectrum~\cite{Friedli:2013id}. As shown schematically in Fig.\,\ref{fig:fwm}\textbf{a}, a single mode DFB QCL and the output of a tunable source based on difference frequency generation were injected in an anti-reflection coated QCL amplifier, driven in continuous wave at room temperature close to its maximum current. The QCL amplifier consisted of a single stack of strain-compensated InGaAs/AlInAs active region operating at 4.3$\mu$m wavelength~\cite{Hinkov:2012iz}. For these operation conditions, the computed upper state lifetime is $T_1 = 0.29$\,ps and dephasing time is $T_2 = 0.14$\,ps. As shown in Fig.\,\ref{fig:fwm} \textbf{b}, the output showed the expected mixing product at frequencies corresponding to $\omega_{4} = 2 \omega_1 - \omega_2$ and $\omega_{3} = 2 \omega_2 - \omega_1$. As shown in Fig.\,\ref{fig:fwm}\textbf{c}, the dependence of FWM signals as a function of detuning was in good agreement with the value of the effective nonlinear susceptibility $\chi^{(3)}$ computed using a two-level density matrix model and given by:
\begin{equation}
\begin{aligned}
\chi^{(3)} (\Delta, \delta \omega) = \frac{2 e^4 \Delta N z_{ij}^4}{3 \epsilon_0 \hbar^3}\frac{(\delta \omega - \Delta - i/T_2)}{(\Delta - \delta \omega + i/T_2)} \\
\times \frac{ (-\delta \omega + 2 i/T_2) (\Delta + i/T_2)^{-1}}{ (\delta \omega - i /T_1)(\delta \omega - \Delta - i T_2)(\delta \omega + \Delta - i/T_2)}
\end{aligned}
\label{eq:chi3}
\end{equation}
where $\Delta = \omega_1 - \omega_{12}$ is the detuning between the pump and the intersubband transition, $\delta \omega = \omega_2 - \omega_1$ the separation between the pump and probe, $z_{ij} = 1.6$\,nm the dipole matrix element and $\Delta N$ the population inversion at threshold. For a detuning $\Delta$ of 210\,GHz, an effective $\chi^{(3)} = 2.5 \times 10^{-15}$\,$\textrm{m}^2\textrm{V}^{-2}$ is predicted, close to the measured value of $\chi^{(3)} = (0.9\pm0.2) \times 10^{-15}$\,$\textrm{m}^2\textrm{V}^{-2}$.
The important conclusion of these measurements was that the QCL active region presents a {\em large, resonant $\chi^{(3)}$}, arising from the {\em active medium itself}. In addition, the bandwidth of the FWM process is much wider than the one in interband semiconductor lasers because of the much shorter upper state lifetime. 

\paragraph*{Waveguide dispersion}
The ability of four wave mixing to generate mode proliferation - and ultimatly comb operation - depends critically on the group velocity dispersion of the cavity. If one considers a single active region with a gain bandwidth of 100\,cm$^{-1}$, the latter corresponds to about $N_m = 200$ cavity modes for a 3\,mm long device. The finesse $\mathcal{F}$ of a typical QCL Fabry-P\'erot cavity is very limited, but even assuming a value of $\mathcal{F} = 5$ requires that the fractional change of the group effective index is not larger than $\delta (\omega)/\omega = \delta n_g/n_g = 1/(N_m \mathcal{F}) = 10^{-3}$ over the gain bandwidth. The spectral change of the group refractive index is quantified by the group velocity dispersion (GVD), defined as
\begin{equation}
\text{GVD} = \frac{\partial }{\partial \omega} \frac{1}{v_g} = \frac{1}{c} \frac{\partial n_g}{\partial \omega}.
\end{equation}
Using the numerical values  reported  above, the GVD must remain below $\text{GVD} < 560$\,fs$^2$/mm for the modes to remain efficiently coupled by four wave mixing. This number is of course only a rough estimate yielding the order of magnitude of the dispersion that is relevant for comb operation.  

The GVD in an active QCL consists of three main components: the dispersion of the material, the modal dispersion of the waveguide (including both lateral and vertical confinement) and the gain medium itself. We have then
\begin{equation}
\text{GVD} = \text{GVD}_{\textrm{mat}}  + \text{GVD}_{\textrm{mod}}  + \text{GVD}_{\textrm{gain}}
\end{equation} 
One very favorable feature of the mid-infrared QCL is the fact that they operate in a transparency region far from both the fundamental gap as well as from the restrahlen band.
\begin{figure}[!htb]
	\includegraphics[width=0.4\textwidth]{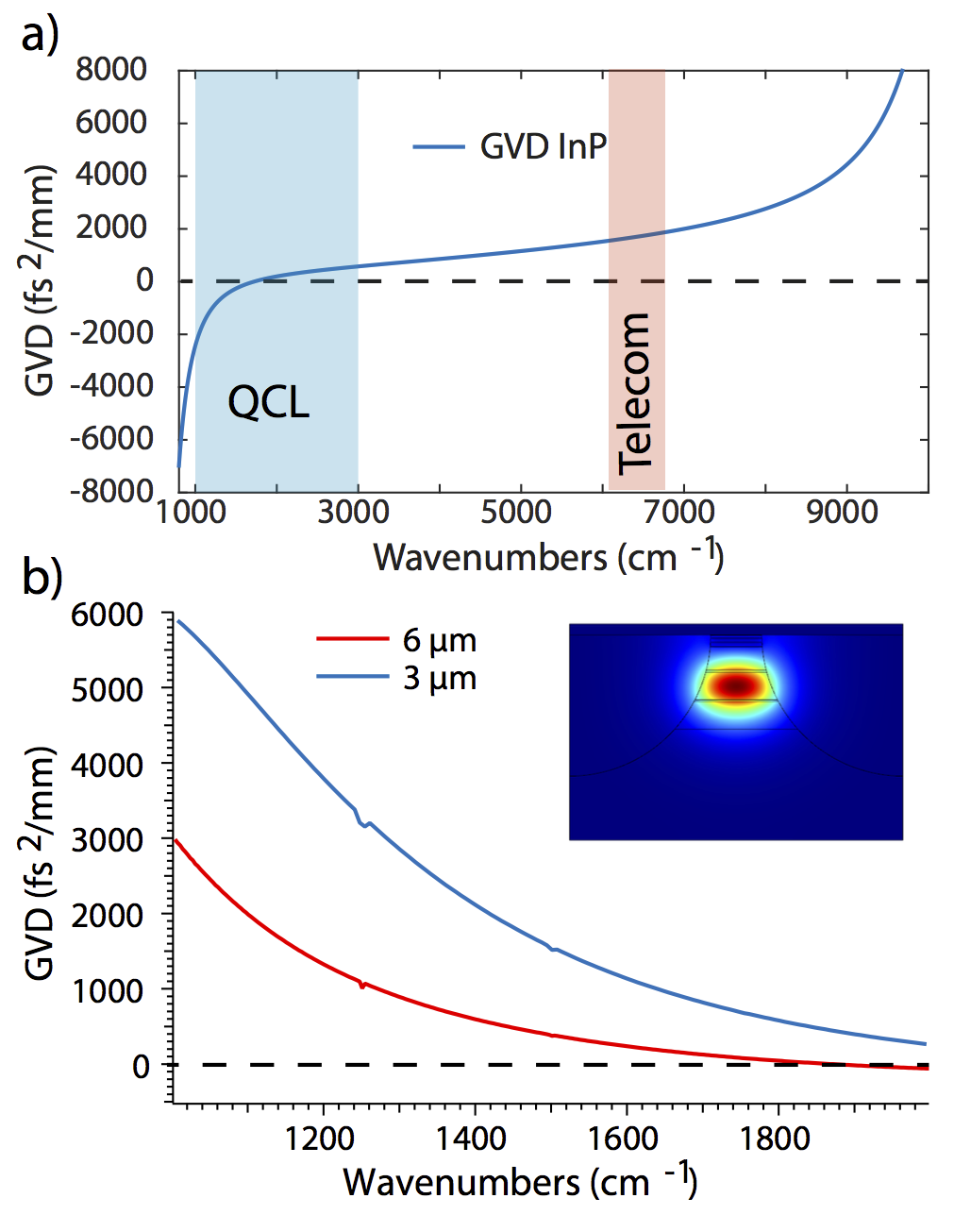}
	\caption{Contributions to the waveguide dispersion. 
	    \textbf{a} GVD for bulk InP material. 
		\textbf{b} GVD due to the mode confinement, computed by performing a two-dimensional simulation, showing the impact of the ridge width on the GVD. Inset: mode profile of a typical mid-infrared QCL.}
	\label{fig:gvd-waveguide}
\end{figure}
For this reason, as shown in Fig.\,\ref{fig:gvd-waveguide}\textbf{a}, the GVD of InP has a very low value ($<1000 \text{ fs}^2/\text{mm}$ in the region covering the mid-infrared range from 3.3-10\,$\mu$m $(\approx 1000-3000$\,cm$^{-1})$. Similar low values are expected for the undoped InGaAs and AlInAs materials constituting the active region. In contrast, the situation at the telecom wavelength of 1.55\,$\mu$m is that InP has already a $\text{GVD}_{\textrm{mat}} \approx  2000$\,$\text{fs}^{2}/\text{mm}$ while the narrower gap confinement waveguide layers will have even larger values of  $\text{GVD}_{\textrm{mat}}$. 

The dispersion of the resulting "empty" waveguide, taking into account the effect of vertical and horizontal confinement is displayed in Fig.\,\ref{fig:gvd-waveguide}\textbf{b}. In that figure, the vertical waveguide geometry was the one of the original work reported~\cite{Hugi:2012ep}. Not surprisingly, strong lateral confinement is shown to add a positive contribution to the GVD, and the choice of the final waveguide width will be the result of a compromise between the need to keep the total GVD low and the necessity to guarantee a single transverse mode emission.  Note also that the original report in~\cite{Hugi:2012ep} the waveguide GVD was computed combining the GVD from the horizontal and vertical confinements separately. We found that the latter procedure was a relatively poor approximation for tightly confined waveguides.  

\begin{figure}[!htb]
	\includegraphics[width=0.45\textwidth]{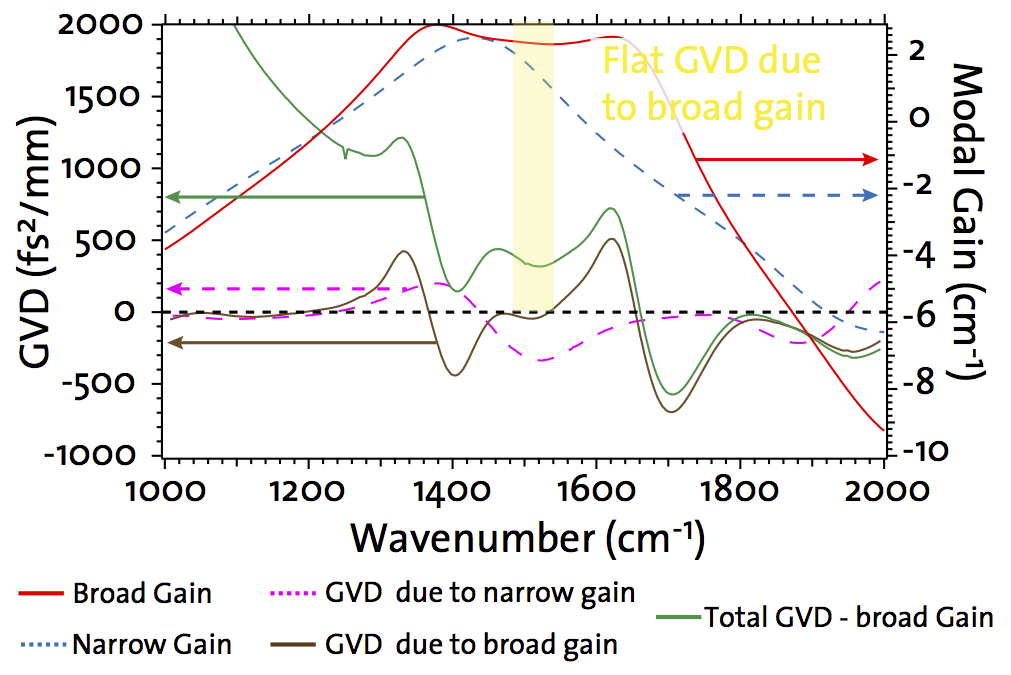}
	\caption{Computation of the GVD for the active region and effect of the gain bandwidth. Blue dashed line: a narrowband single stack active region. Solid red line: active region consisting of three stack. Results published in~\cite{Hugi:2012ep}.}
	\label{fig:gvd-total}
\end{figure}
Finally, the gain itself is responsible for a group velocity dispersion because of the change of refractive index following Kramers-Kr\"onig relations. Shown in Fig.\,\ref{fig:gvd-total} is the computation of the complete QCL, including gain, comparing the case of the three stack active region device (in solid line) with a hypothetical single stack (dashed line). The parameters are the ones of the broadband device exhibiting comb operation~\cite{Hugi:2012ep}. At maximum power, the device is operating with a continuous spectrum covering 250\,cm$^{-1}$, implying that the net gain is equalized over this bandwidth. For this reason, the GVD originating from the gain exhibits a wide flat minimum in the middle of the gain curve, in contrast to the case of the single stack QCL. In this simulation, the total GVD (solid black line) was predicted to exhibit a relatively wide flat minimum in the region around 1450\,cm$^{-1}$, in good agreement with our observation of comb operation in the center of the gain curve.

One difficulty arising from the computation done here is that the shape of the GVD depends strongly on the shape of the gain curve, which itself is determined self-consistently by gain saturation under operation. This special situation arises because, as compared to the case of solid-state or fiber lasers, the value of gain in QCL is large such that dispersion and gain cannot be separated easily. Treating both contributions on the same footing is one of the nice feature of the Maxwell-Bloch model presented in the next paragraph.

\paragraph*{Theoretical models for comb operation}
The dynamical behavior of a multimode QCL laser is the result of a complex interplay between the intersubband polarization in the active region and the electromagnetic energy stored in the optical modes of the laser cavity. The equations of motion for the carriers in the active region can be represented by the optical Bloch equations for the density matrix describing the laser transition as a two-level system.  The Maxwell equations describe then the motion of the electromagnetic wave subjected to the matter polarization and losses in the cavity. The resulting Maxwell-Bloch coupled equations are the basis for the explanation of laser dynamics. 
\begin{figure}[!htb]
	\centering
	\includegraphics[width=0.3\textwidth]{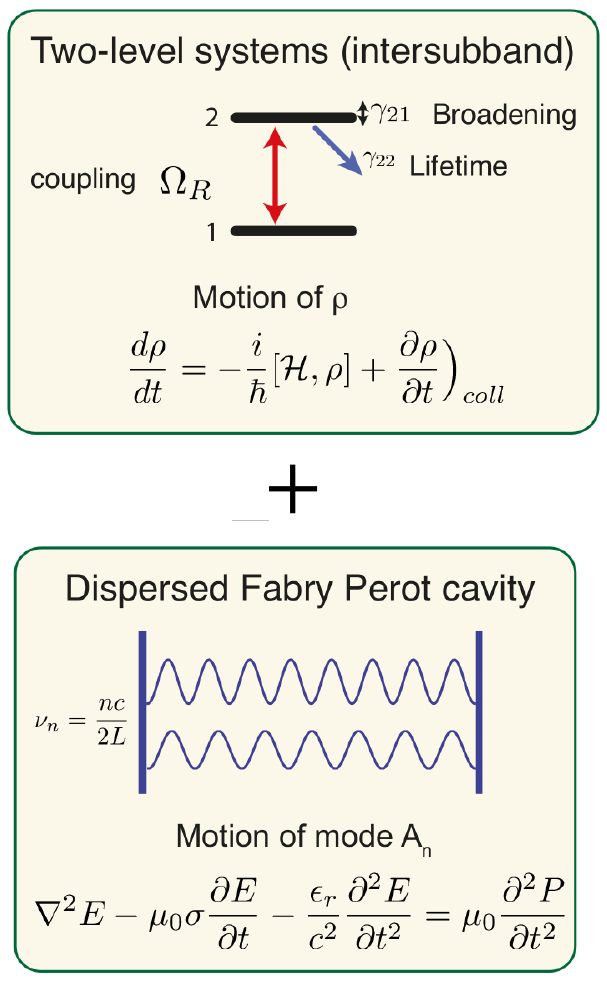}
	\caption{Schematic drawing showing the coupling between the optical Bloch equations, describing the motion of the density matrix and the Maxwell equation for the optical modes.}
	\label{fig:maxwell-bloch}
\end{figure}

Narrowband modulation at the round trip frequency of the output power of continuous wave QCLs were first interpreted using a set of Maxwell-Bloch equation expressed in the time domain~\cite{Wang:2007p232,Gordon:2008p310,Gkortsas:2010tu}. However, these equations, while well suited to describe the generation and propagation of a pulse in the laser cavity, are ill-conditioned to study the output of a laser that exhibits mostly a frequency modulated output. 

\begin{figure}[!htb]
	\centering
	\includegraphics[width=0.45\textwidth]{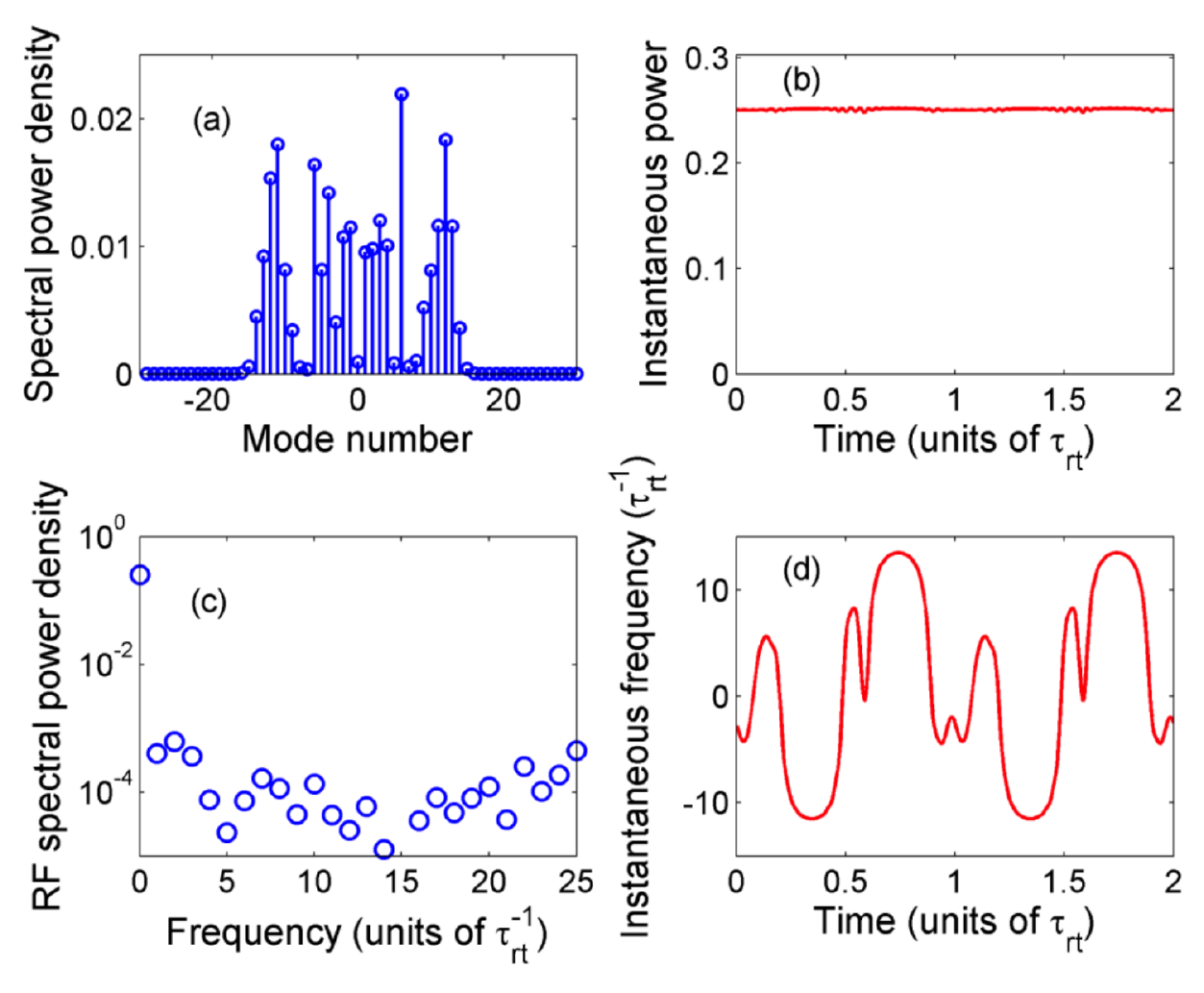}
	\caption{ Maxwell-Bloch model of a  mid-infrared QCL comb. 
		\textbf{a} Optical spectrum. 
		\textbf{b} Instantaneous intensity as a function of time.  
		\textbf{c} RF spectrum of the intensity. 
		\textbf{d} Instantaneous optical frequency. Results published in~\cite{Khurgin:2014hy}.}
	\label{fig:comb-Jakob}
\end{figure}

Another approach consists solving the system in the frequency domain, using the modes of the laser cavity as a basis. This approach was already used by Lamb in 1964 to study the "three-mode laser", in which a third mode is frequency locked by the beating of two other modes through the four wave mixing~\cite{LambJr:1964td}. 
The method followed by Khurgin et al.~\cite{Khurgin:2014hy} uses this approach and applies it to an arbitrary set of modes $N$ as shown schematically in Fig.\,\ref{fig:maxwell-bloch}. The final equation for the time evolution of the (slowly varying and normalized) complex amplitude $A_n$ of mode $n$ is given by ~\cite{Khurgin:2014hy,Villares:2015ho}:
%
\begin{eqnarray}
 \dot{A_n} = \left \{ \underbrace{G_n - 1}_{\hbox{\scriptsize{Net gain}}} +  i  \underbrace{\left( \frac{ \omega_n^2 - \omega_{nc}^2 }{ 2 \omega_n} \right )}_{\hbox{\scriptsize{Cavity dispersion}}} \right \} A_n  
  \nonumber \\
  -   G_n \sum_{k,l}  
 \underbrace{ 
  C_{kl} B_{kl} A_{m} A_{l} A_{k}^* \kappa_{n,k,l,m}    
}_{\hbox{\scriptsize{FWM term}}}  
. 
\label{eq:modedynamicsall}
\end{eqnarray}
In the above equation, $G_n$ is the transition net gain, normalized to the cavity losses:
 \begin{equation}
G_n = \frac{1}{1- i n \omega/\gamma_{12}}
\end{equation}
while
 \begin{equation}
 C_{kl} = \frac{ \gamma_{22}}{\gamma_{22} - i (l-k) \omega}
 \end{equation}
  defines the amplitude of the coherent population oscillations, and  
\begin{equation} 
B_{kl} = \frac{\gamma_{12} }{2 i} \left (  \frac{1}{-i \gamma_{12} - l \omega}  - \frac{1}{i \gamma_{12} - k \omega} \right ) 
\end{equation}
is the term driving the width of the four wave mixing gain. In these equations, $\omega_{nc}$ is the angular frequency of the empty cavity mode, $\omega_n = n \omega$  is the angular frequency of one specific mode $n$. The broadening of the transition is $\gamma_{12}$ and the scattering rate of the upper state is $\gamma_{22}$. The result of the simulation, neglecting at this stage the effect of the dispersion, is reported in Fig.\,\ref{fig:comb-Jakob}. The results show that the short upper state lifetime prevent the formation of optical pulses; as a result the instantaneous intensity is almost constant (Fig.\,\ref{fig:comb-Jakob}\textbf{b}) while the instantaneous frequency swings widely over a significant portion of the output spectrum (Fig.\,\ref{fig:comb-Jakob}\textbf{d}). Note that the optical Bloch equations used in the derivation of Eq.~\ref{eq:modedynamicsall} are the same as the one used for the derivation of the $\chi^{(3)}$ in Eq.~\ref{eq:chi3}; resulting in the expression (in normalized units) $\chi^{(3)} = G_n \sum_{k,l} B_{kl} C_{kl}$.

In insight, these results are to be expected since QCL are so-called class A laser, where the medium population inversion and polarization can follow the optical field dynamics. Indeed, FM mode-locking was observed in He-Ne gas lasers, that belong to the same class A~\cite{HARRIS:1965up}. In that experiment, however, the FM mode-locking was driven by an external phase modulator. Interestingly, it was argued on theoretical grounds that self mode-locking through four wave mixing was not possible. However, it seems that the author did not consider the possibility of a FM-like solution~\cite{Khanin:1995wt}. 

\begin{figure}[!htb]
	\centering
	\includegraphics[width=0.45\textwidth]{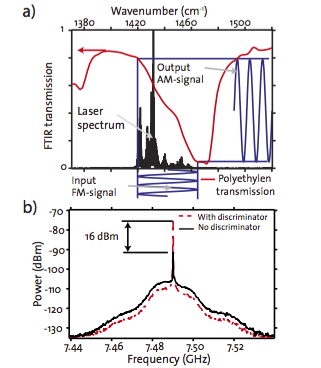}
	\caption{ Frequency-modulated QCL comb. 
		\textbf{a} Optical spectrum (black) as well as polyethylen transmission spectrum (red) acting as an optical discriminator.
		\textbf{b} RF spectrum showing the beatnote at the round trip frequency. An increase of the RF beatnote, characteristics of an FM to AM conversion.}
	\label{fig:polyethylen_discriminator}
\end{figure}

Figure Fig.\,\ref{fig:polyethylen_discriminator} shows an experimental confirmation of the FM-like beahavior of a QCL comb. A sheet of polyethylene, which
has strong wavelength-dependent absorption (see red line in Fig.\,\ref{fig:polyethylen_discriminator}\textbf{a}), is inserted between the laser output and a high-bandwidth detector. This sheet acts as an optical discriminator that can convert the frequency modulation of the laser output to an amplitude modulated signal. We observe an amplification (16\,dB in this example) of the measured RF beatnote at the round trip frequency (see Fig.\,\ref{fig:polyethylen_discriminator}\textbf{b}), as expected from a frequency-modulated signal.

The effect of cavity dispersion and external modulation on the solutions of Eq.~\ref{eq:modedynamicsall} was studied in ref.~\cite{Villares:2015ho}. As expected, a large absolute value of GVD (= 50'000\,fs$^2$/mm) prevented the formation of the comb. However, at this point the simulations could not conclusively assess the role of small values of GVD ($\approx 500$\,fs$^2$/mm) in the destabilisation of the coherence. Again, as expected, the same model could predict the formation of single pulses by modulation of a section of the waveguide in THz devices with long upper state lifetime~\cite{Villares:2015ho}, essentially reproducing the experimental results observed in ~\cite{barbieri:2011p1986}. 

However, the role of electron transport is not sufficiently taken into account in the existing models, although it has been recently included in the description of actively mode-locked QCLs~\cite{Wang:2015gj}. In particular, some experimental evidence show that embedding the QCL comb in a RF microstrip waveguide improves the coherence of the beatnote~\cite{StJean:2014hl} while this effect is not predicted by the computer simulations~\cite{Villares:2015ho}. A more complete theory would incorporate the second (slower) time scale of the injection process as well as a $\chi^{(2)}$ interaction where a mode is shifted in energy by the interaction with the RF modulation at the round trip frequency. 

\paragraph*{Comb operation of Mid-IR broadband QCL}

\begin{figure}[!htb]
	\includegraphics[width=0.45\textwidth]{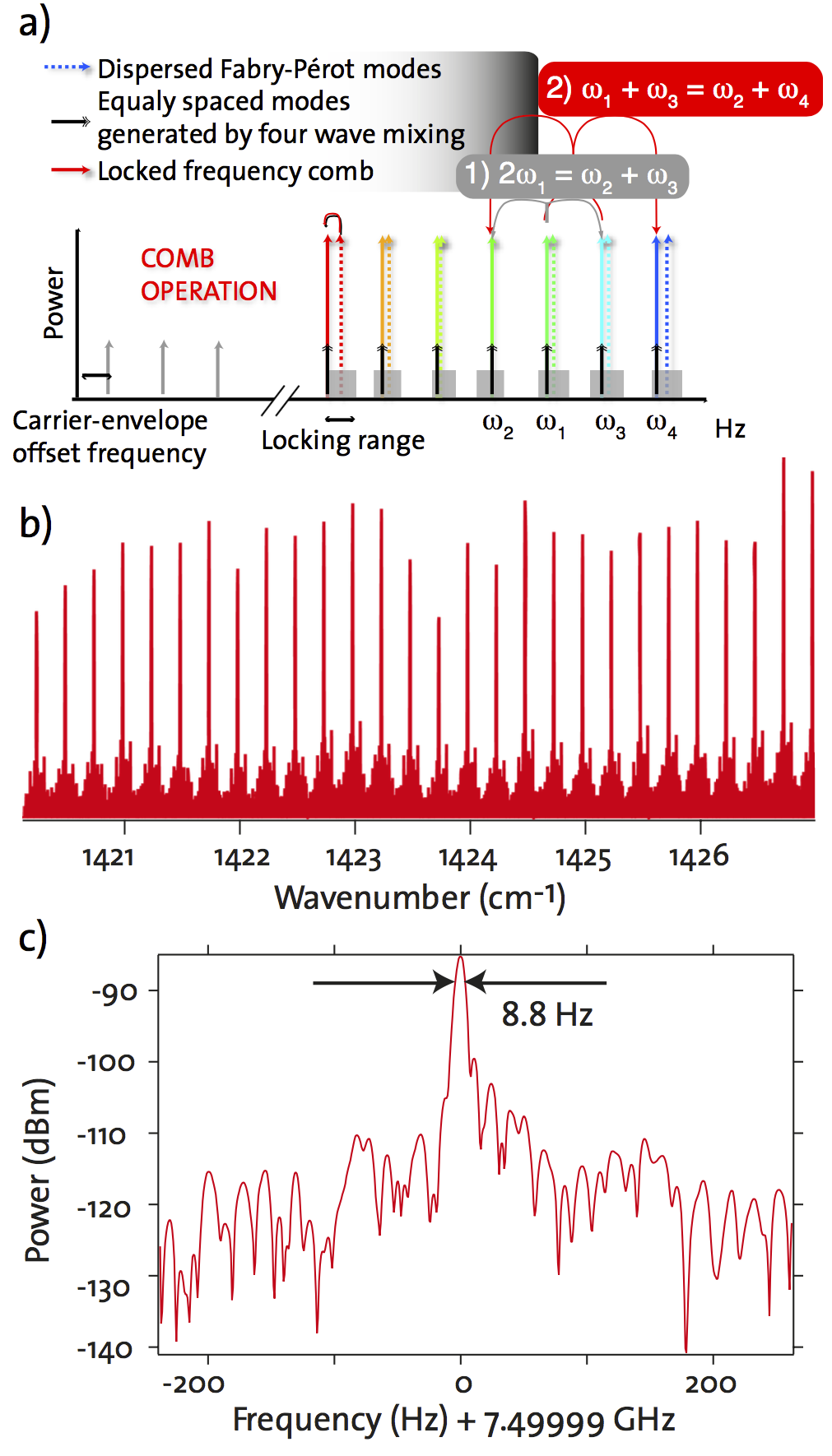}
	\caption{
		Principle of operation of a QCL comb:
		\textbf{a} The dispersed Fabry–P\'erot resonator modes are injection-locked by the equally spaced modes generated by four-wave mixing processes.
		\textbf{b}. Magnified view of the optical spectrum of a QCL comb operating in the mid-infrared, at a center wavenumber of 1430\,cm$^{-1}$ (7\,$\mu$m) covering 60\,cm$^{-1}$.
		\textbf{c}. Radio-frequency spectrum of the intensity near the repetition frequency $f_{\textrm{rep}}$, showing a beatnote with a full-width at half-maximum of $<10$\,Hz. The measurement is taken at the onset of the multimode emission and the resolution bandwidth of the spectrum analyser is set to 10\,Hz. Results published in~\cite{Hugi:2012ep}.
	}
	\label{fig:QCL_comb_operation}
\end{figure}

Frequency combs are generated when the different longitudinal modes of a laser are locked in phase~\cite{Udem:2002p2021,Diddams:2010p2046}, creating an array of equidistantly spaced, phase-coherent modes. In contrast with previous attempts of mode-locking of QCLs~\cite{Wang:2009p1554,Barbieri:2010p1763,barbieri:2011p1986,Ravaro:ij}, broadband QCLs can achieve frequency comb operation by using four-wave-mixing as a phase-locking mechanism~\cite{Hugi:2012ep}, as shown schematically in Fig.\,\ref{fig:QCL_comb_operation}\textbf{a}. Degenerate and non-degenerate four-wave mixing processes induce a proliferation of modes over the entire laser spectrum. The dispersed Fabry-P\'erot modes -- which are present in the case of a free-running multimode laser and are not phase-locked -- can be injection-locked by the modes generated by the FWM process, thus creating a frequency comb. This phase-locking mechanism is very similar to the one that occurs in microresonator-combs~\cite{Delhaye:2007p1571,Kippenberg:2011p2042}. In addition, combined with the short gain recovery of intersubband transitions, the laser output is not pulsed and resembles the one of a frequency modulated laser~\cite{Hugi:2012ep}. In fact, the fast gain recovery of a QCL does not introduce any perturbation to a frequency-modulated laser, as the intensity of such a laser is constant.

Fig.\,\ref{fig:QCL_comb_operation}\textbf{b} shows a magnified view of the optical spectrum of a QCL comb operating in the mid-infrared, at a center wavenumber of 
1430\,cm$^{-1}$ (7\,$\mu$m) covering 60\,cm$^{-1}$. We observe a single set of longitudinal modes with no measurable dispersion within the resolution of the measurement (0.0026\,cm$^{-1}$, 78\,MHz). As this measurement does not give any information about the relative stability of the phases of the modes, different techniques had to be developed in order to characterize QCL combs (see next section).

In the first place, the radio-frequency spectrum of the laser intensity near the repetition frequency $f_{\textrm{rep}}$ can be measured. High-bandwidth detectors capable of detecting radio-frequency signals up to tens of GHz are needed for this type of characterization. Such a spectrum would show a beatnote with a large linewidth (5 to 10\,MHz) in the case of two uncouples modes (such as those generated by two independent lasers). In contrast, such a measurement would show a narrow beatnote in case the modes are locked in phase. The measurement of this radio-frequency spectrum, also called intermode beatnote spectrum, is shown in Fig.\,\ref{fig:QCL_comb_operation}\textbf{c}, for a laser biased just above the onset of multimode operation. It shows a beatnote with a linewidth with a full-width at half-maximum of 8.8 Hz, characteristic of comb operation.

\begin{figure}[!htb]
	\includegraphics[width=0.5\textwidth]{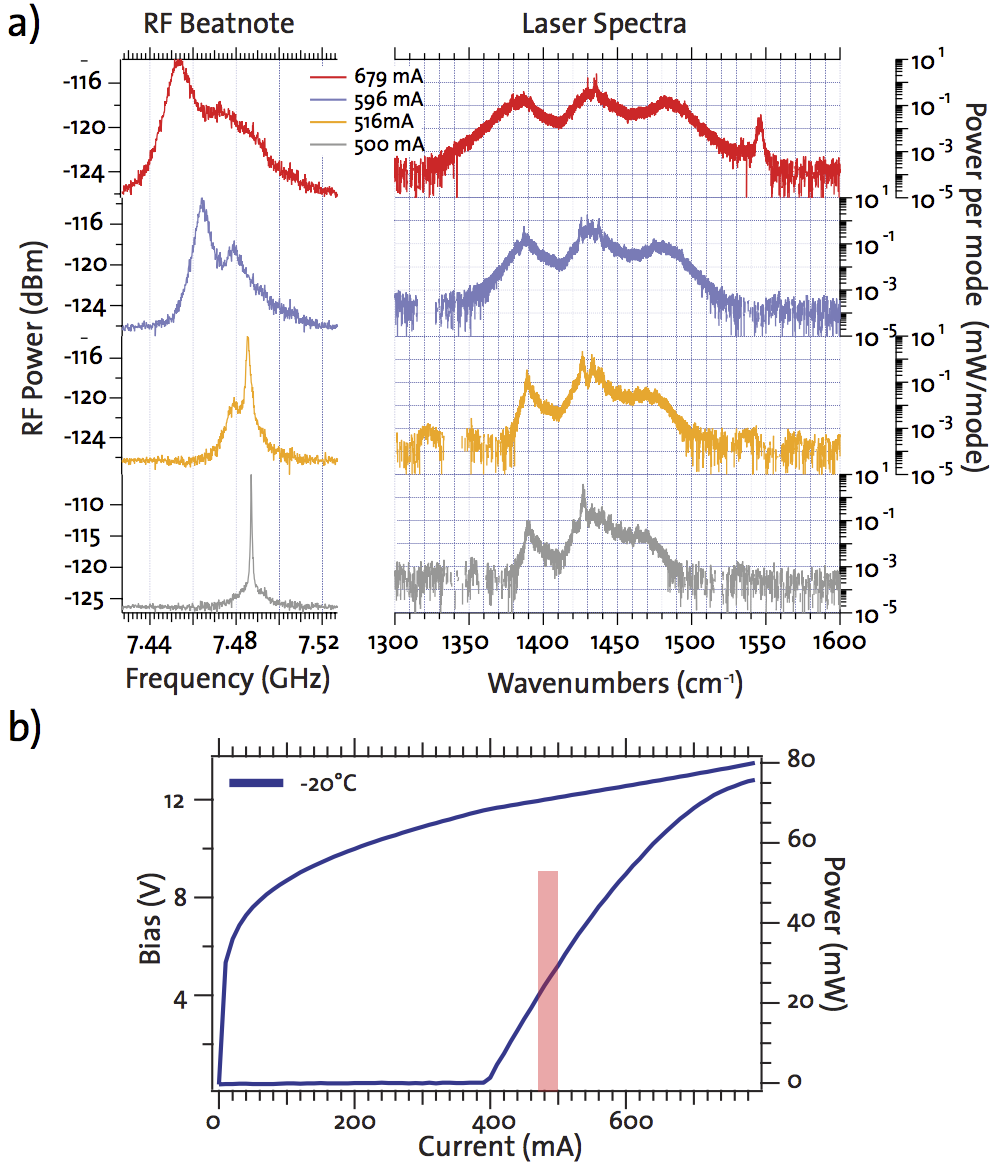}
	\caption{
		QCL comb operation regimes.
		\textbf{a} Left side: RF beatnote spectra measured using a fast quantum well photoconductor, together with its respective optical spectra (right graphs) of a QCL designed for comb operation. Above a certain value of current (510\,mA in this example), the RF beatnote splits into two distinctive beats. 
		\textbf{b)} Corresponding light intensity-current-voltage characteristics of the device. The shaded area corresponds to the region where the narrow beatnote is observed. Results published in~\cite{Hugi:2012ep}. 
	}
	\label{fig:QCL_comb_regimes}
\end{figure}

Frequency comb operation was however not observed over the entire laser operation range in the original work of ref. ~\cite{Hugi:2012ep} and these observations were also confirmed in QCL combs operating in the THz spectral range~\cite{Burghoff:NATPHOT:2014,Rosch:2014ft}. Fig.\,\ref{fig:QCL_comb_regimes}\textbf{a} shows different RF spectra together with their corresponding optical spectra for different values of laser current. In addition to the narrow RF beatnote characteristic of comb operation, broad beatnotes are also observed, corresponding to a loss of coherence of the optical comb. Fig.\,\ref{fig:QCL_comb_regimes}\textbf{b} shows a light intensity-current-voltage characteristics of this same device, showing the range where QCL comb operation is observed.  

\begin{figure}[!htb]
	\includegraphics[width=0.45\textwidth]{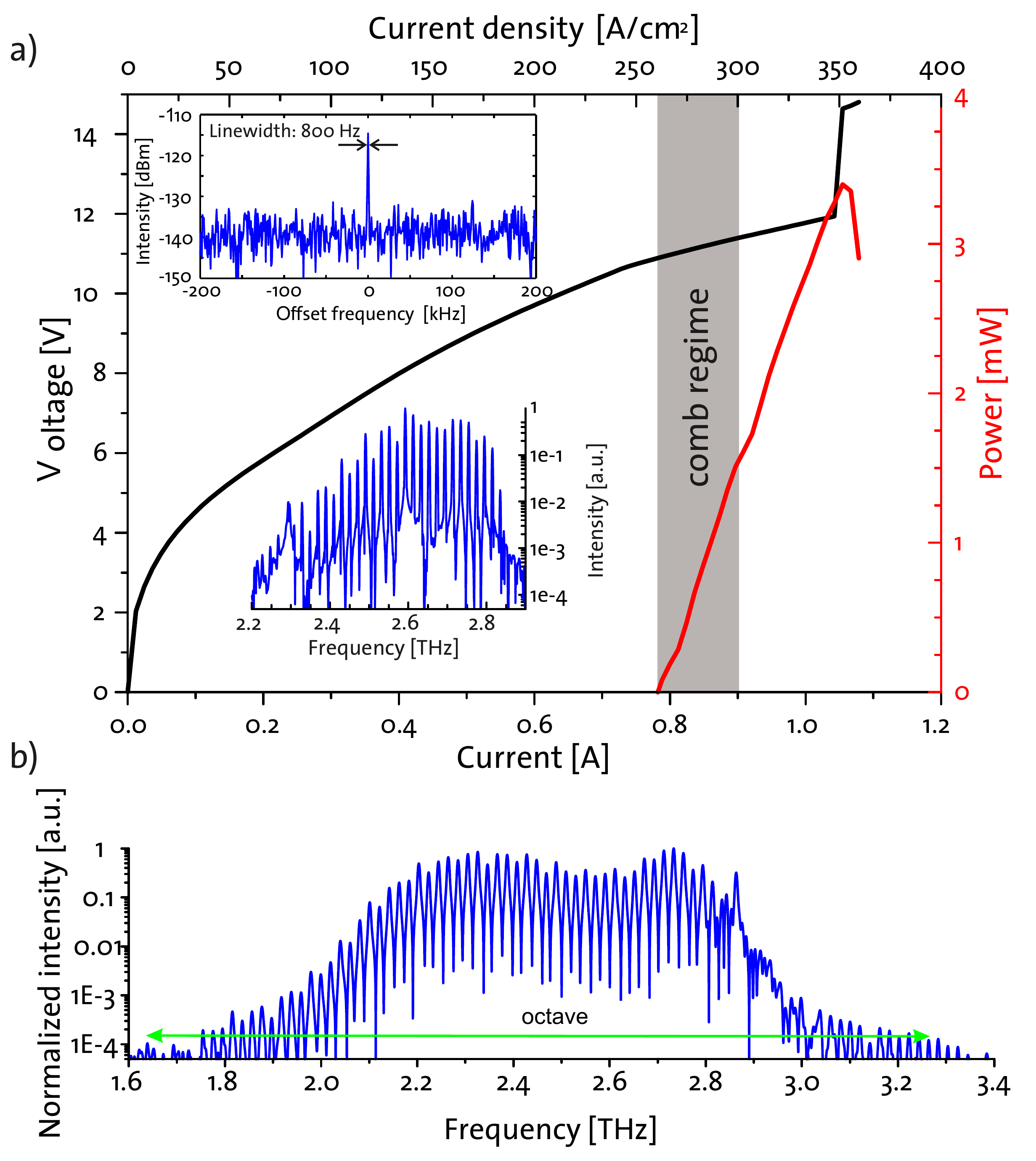}
	\caption{
		THz QCL comb.
		\textbf{a} LIV curves of a 2mm x 150\,$\mu$m long laser measured at 25 Kelvin in CW operation. The shaded area indicates the current range where the laser operates as a comb. The upper inset displays the intermode beatnote, showing a linewidth of 800\,Hz at a driving current of 0.9\,A. The lower inset shows the corresponding optical spectrum with a bandwidth of 507\,GHz. 
		\textbf{b} Octave-spanning spectrum of the same laser measured at 25 Kelvin in CW operation at the maximum current  (1.05\,A). Results published in~\cite{Rosch:2014ft}.
		}
	\label{fig:THz_Octave_Spanning}
\end{figure}

\begin{figure}[!htb]
	\includegraphics[width=0.45\textwidth]{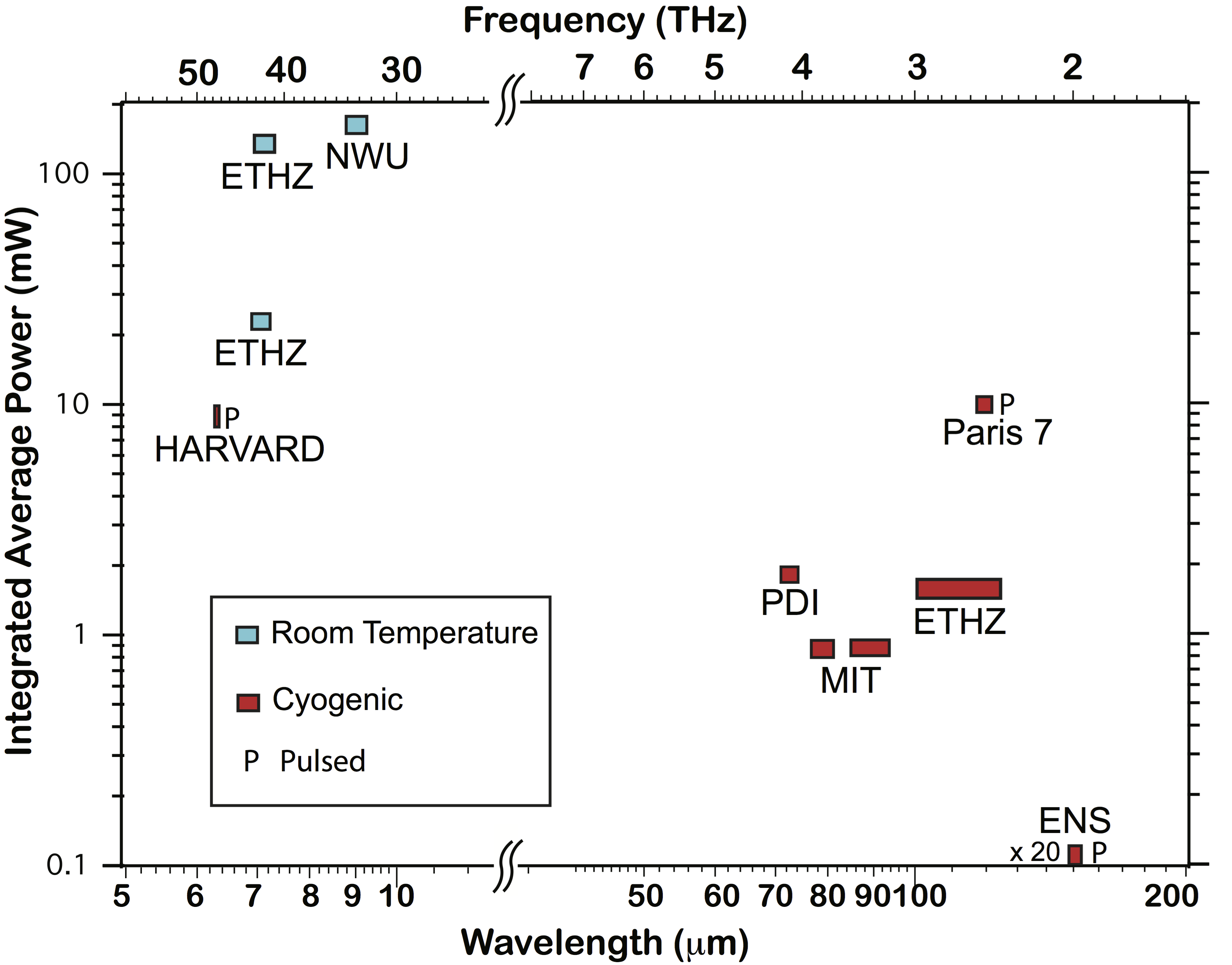}
	\caption{Average emitted power as a function of  frequency coverage for published QCL-based comb sources. Rectangles width indicates the frequency coverage and the average power refers to the integrated laser output on all the emitted wavelengths. Data from the following groups: ETHZ \cite{Hugi:2012ep,Rosch:2014ft,Villares:2015ti}, NWU \cite{Lu:2015en}, Paris 7 \cite{barbieri:2011p1986}, PDI \cite{Wienold:2014fb}, Harvard \cite{Wang:2009p1554}, ENS \cite{Freeman:2012hd}, MIT\cite{Burghoff:NATPHOT:2014}.
	}
	\label{fig:Comb_power}
\end{figure}

\paragraph*{THz QCL  comb operation}	
In addition to mid-infrared combs, QCLs can be used to generate combs at THz frequencies, as the main driving mechanism is still four-wave mixing. THz QCLs can be driven in the comb regime by careful dispersion compensation~\cite{Burghoff:NATPHOT:2014}, by engineering a broadband QCL with a flat gain curve and by a combination of both elements.
It was recently shown that it is possible to achieve octave-spanning lasers at THz frequencies using QCLs~\cite{Rosch:2014ft}. This very wide frequency coverage is possible due to the gain engineering capability typical of intersubband transitions coupled to the extremely broadband nature of the double-metal waveguide. THz QCLs operate in double-metal cavities where the waveguide claddings are constituted by two metallic layers \cite{Scalari:2009p746}. In TM polarization this cavity does not present any cutoff, making ultra-broadband operation possible. The heterogenous cascade laser used for achieving octave-spanning operation is constituted by three different active regions stacked together in the same waveguide. Its emission spans from 1.64\,THz to 3.35\,THz when operated in CW operation at a temperature of 25 Kelvin (Fig.\,\ref{fig:THz_Octave_Spanning}b) \cite{Rosch:2014ft}. Such a broad gain medium is beneficial for comb operation as it results in a low and flat GVD. Furthermore the  double metal waveguides introduce only a small amount of GVD to the laser. As a result, these lasers can act as frequency combs similar as mid-infrared QCL combs. The characteristics of such a THz QCL comb are displayed in Fig.\,\ref{fig:THz_Octave_Spanning}a. The comb shown has 1.55\,mW of output power while spanning over a spectral bandwidth of 507\,GHz. The beatnote exhibits a linewidth of 800\,Hz limited by technical noise. So far comb operation in these devices is limited to a spectral bandwidth of the order of 600\,GHz at a central frequency of 2.6\,THz. Additionally these combs suffer from the general drawbacks of THz QCLs: operation is limited to cryogenic temperatures and feature much lower output power than combs in the mid-infrared (of the order of 10-100\,mW).  \\
Fig.\,\ref{fig:Comb_power} gives an overview over the output powers of different QCL-based frequency combs. The power levels range from 1-10\,mW for THz combs up to more than 100\,mW for mid-infrared combs. As can be seen from the same figure, the frequency coverage of the different combs show large differences. 

\section{Comb coherence}
When discussing the coherence properties of a frequency comb, two different quantities have to be distinguished. The \emph{relative coherence} between the different comb modes describes the relative phase stability between any two tooths of a comb. This coherence property is intrinsically linked to the comb repetition frequency $f_{\textrm{rep}}$, and a frequency comb with high degree of relative coherence shows perfectly spaced modes. On the other hand, the \emph{absolute coherence} of a frequency comb characterizes the global phase stability of all modes. This second coherence property is linked to the comb offset frequency $f_{\textrm{ceo}}$.   

\begin{figure*}[!htb]
	\includegraphics[width=1\textwidth]{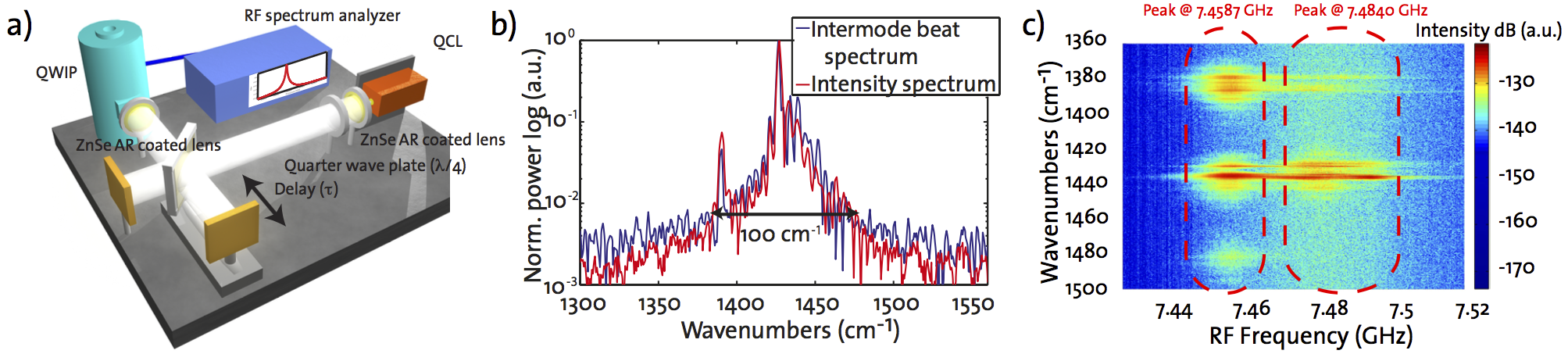}
	\caption{Intermode beat spectroscopy.
		\textbf{a} Schematic representation of an intermode beat spectroscopy experiment. The output of a FTIR is sent to a high-bandwidth detector, in order to measure the intermode beatnote spectrum with a spectrum analyzer. (QWIP: quantum-well infrared photodetector. AR: anti-reflection). 
		\textbf{b} Intermode beat spectrum together with the normal intensity spectrum of the laser, showing that the entire spectrum is contributing to the generation of the beatnote at the comb repetition frequency.
		\textbf{c} Intermode beat spectrum acquire over a large span ($100$\,MHz), for driving conditions where the QCL is not operating as a comb, and yielding important informations about the comb destabilization mechanisms. Two broad beatnotes are observed on the RF spectrum. The peak at $\nu = $ 7.4840\,GHz involves mostly modes which are near the center of the optical spectrum ($\omega \simeq 1430$ cm$^{-1}$) while the peak at $\nu = $ 7.4587\,GHz corresponds to modes locked for a wider bandwidth (from $\omega=1360$ cm$^{-1}$ to $\omega=1500$ cm$^{-1}$). Results published in~\cite{Hugi:2012ep}. 
	}
	\label{fig:intemodebeat_spectroscopy}
\end{figure*}
As QCL combs are based on a different phase-locking mechanism as compared to frequency combs emitting pulses, new characterizations techniques had to be developed to assess their coherence properties. Indeed, traditional characterization techniques -- such as interferometric autocorrelation, FROG or others~\cite{trebino1997measuring} -- are all based on nonlinear effects induced by the short optical pulse emitted by traditional frequency combs and cannot be used for the characterization of QCL combs.

For this reason, characterization techniques which did not require pulse emission were developed. The first technique developed for the characterization of the relative coherence of QCL combs is an interferometric technique where the autocorrelation of the beatnote at the comb repetition frequency is measured using a Fourier Transform Infrared Spectrometer (FTIR)~\cite{Hugi:2012ep}. 

\paragraph*{Intermode beat spectroscopy and SWIFTS}

The principle of \emph{intermode beat spectroscopy} is schematically described in Fig.\,\ref{fig:intemodebeat_spectroscopy}\textbf{a}. The FTIR act as an optical filter used for the spectral separation of the comb modes, and the relative coherence is measured by investigating the beatnote at the comb repetition as a function of the interferometric delay. This is done by using a high-bandwidth detector, sufficiently fast to measure the comb repetition frequency. The quantity measured is given by
\begin{equation}
I_{\textrm{BS}}(\nu,\tau) \equiv \Bigg|\mathcal{F}\bigg\{\big|E(t)+E(t+\tau)\big|^{2}\bigg\}\Bigg|(\nu)
\label{eeq:intermode_beat_spectroscopy}
\end{equation}
which is the absolute value of the Fourier component of the intensity at the detector at frequency $\nu$ over a chosen resolution bandwidth of the spectrum analyser. The quantity $E$ denotes the time-dependent amplitude of the electric field and $\mathcal{F}$ stands for the Fourier transform that is applied to the detected intensity. The traditional intensity interferogram measured in a FTIR is given by $I_{\textrm{BS}}(0,\tau)$. The intermode beat interferogram $I_{\textrm{BS}}(\nu,\tau)$ is sensitive to the relative phase of the modes and can be used to measure the relative coherence of QCL combs. In analogy with Fourier-Transform spectroscopy, one can calculate the Fourier transform of $I_{\textrm{BS}}(\nu,\tau)$ as

\begin{equation}
\mathcal{I}_{\textrm{BS}}(\nu,\omega) \equiv \mathcal{F}(I_{\textrm{BS}}(\nu,\tau))
\label{eeq:intermode_beat_spectroscopy}
\end{equation}

where $\omega$ is usually an optical frequency. The quantity $\mathcal{I}_{\textrm{BS}}(\nu_{rep},\omega)$ evaluated at the comb repetition frequency, called the intermode beat spectrum, is shown in Fig.\,\ref{fig:intemodebeat_spectroscopy}\textbf{b}, together with the normal intensity spectrum of the laser. One can see that both spectrum are very similar, showing that the entire spectrum is contributing to the generation of the beatnote at the comb repetition frequency.

Another important characteristic of intermode beat spectroscopy is that one does not need to necessarily impose $\nu = \nu_{rep}$ and can therefore measure the entire intermode beat spectrum around $\nu_{rep}$. This can be used to yield important information about the mechanisms that destabilize the comb. Fig.\,\ref{fig:intemodebeat_spectroscopy}\textbf{c} shows an example of such characteristics, where the intermode beat spectrum $\mathcal{I}_{\textrm{BS}}(\nu,\omega)$ was acquired over a span of $100$\,MHz around the comp repetition frequency. In this measurement, two broad beatnotes are observed on the RF spectrum. The intermode beat spectrum $\mathcal{I}_{\textrm{BS}}(\nu,\omega)$ shows that the peak at $\nu = $ 7.4840\,GHz involves mostly modes which are near the center of the optical spectrum ($\omega \simeq 1430$\,cm$^{-1}$) while the peak at $\nu = $ 7.4587\,GHz corresponds to modes locked for a wider bandwidth (from $\omega=1360$\,cm$^{-1}$ to $\omega=1500$\,cm$^{-1}$).
     
\begin{figure*}[!htb]
	\includegraphics[width=1\textwidth]{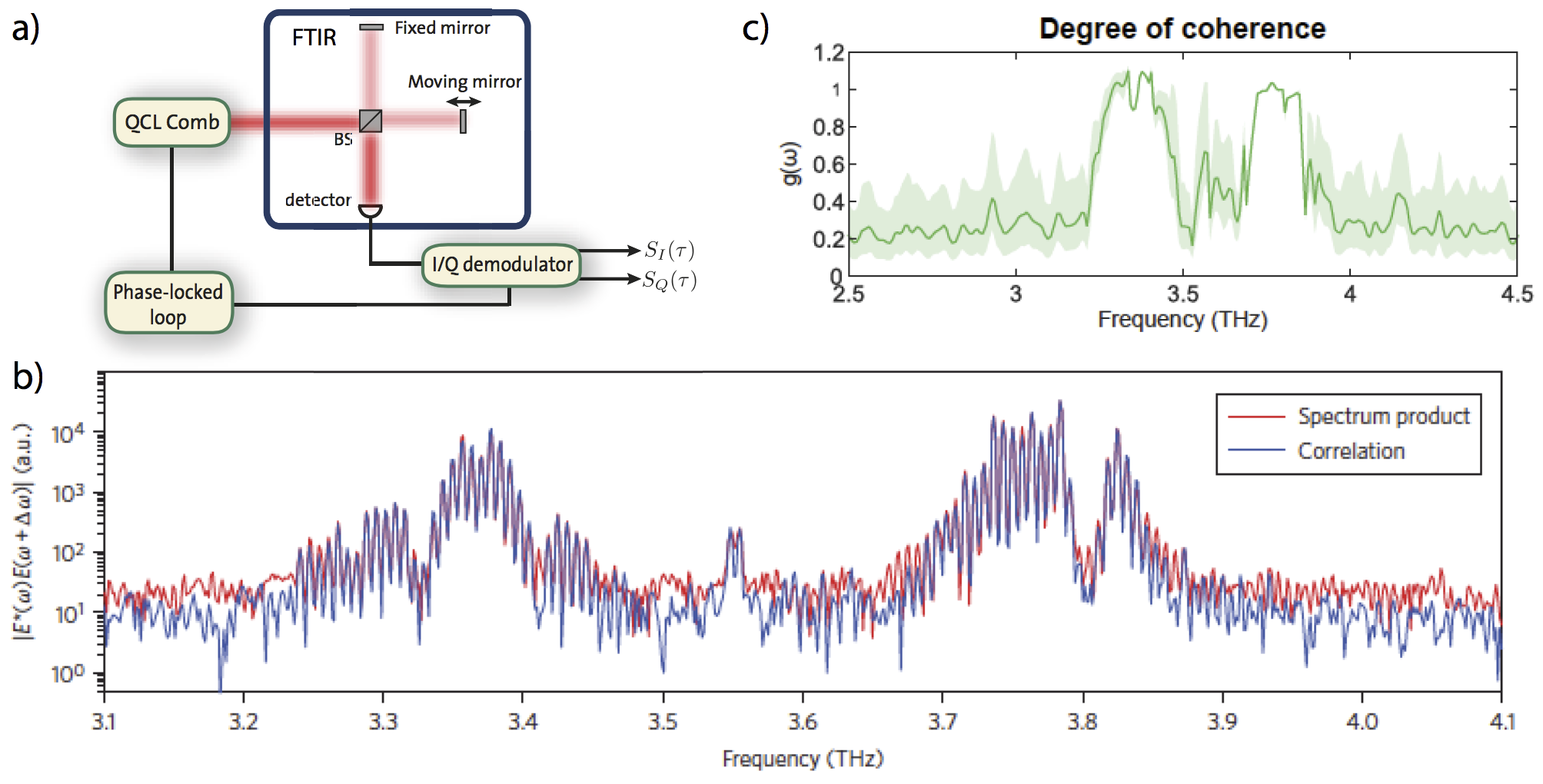}
	\caption{Shifted wave interference Fourier Transform spectroscopy (SWIFTS).
		\textbf{a} Schematic representation of SWIFTS characterization system. The output of a FTIR is sent to a high-bandwidth detector. The detected signal is sent to an I/Q demodulator in order to get access to the in-phase and quadrature signals, $S_{I}(\tau,\omega_{0})$ and $S_{Q}(\tau,\omega_{0})$,respectively. By combining both signals, the correlation of the electric field of two adjacent modes  $<E^{*}(\omega)E(\omega\pm\nu_{\textrm{rep}})>$ can be calculated. The intermode beatnote of the QCL has to be phase-locked to the Local Oscillator of the I/Q Modulator. 
		\textbf{b} SWIFTS correlation spectrum together with the spectrum product. Similar to the intermode beat spectroscopy, these results analysed with SWIFTS also shows that most of the spectral power is in the frequency comb.
		\textbf{c} Degree of relative coherence $g_{\pm}(\omega,\omega\pm\nu_{\textrm{rep}})$ obtained by SWIFTS.
		Results published in~\cite{Burghoff:NATPHOT:2014,burghoff2015evaluating}.} 
	\label{fig:SWIFTS}
\end{figure*}

Although intermode beat spectroscopy can give access to the relative coherence of the modes, it can not give direct access to the relative phases between two modes. For solving this issue, shifted wave interference Fourier Transform Spectroscopy (SWIFTS) has been introduced~\cite{Burghoff:NATPHOT:2014,burghoff2015evaluating}. A schematic description of SWIFTS is shown in Fig.\,\ref{fig:SWIFTS}\textbf{a}. The RF beatnote detected by a high bandwidth detector is now demodulated by a I/Q demodulator, consisting of a local oscillator (LO) creating a pair of signals different by a 90$^{\circ}$ phase shift. By phase-locking the comb repetition frequency to the LO, this measurement yields the in-phase and quadrature signals $S_{I}(\tau,\omega_{0})$ and $S_{Q}(\tau,\omega_{0})$, respectively (see Table \ref{tab:table1} for a comparison between the two methods). By combining both signals, the degree of relative coherence between two modes can be obtained~\cite{burghoff2015evaluating}

\begin{equation}
g_{\pm}(\omega,\omega\pm\nu_{\textrm{rep}}) \equiv \frac{<E^{*}(\omega)E(\omega\pm\nu_{\textrm{rep}})>}{\sqrt{<|E(\omega)|^{2}><|E(\omega\pm\nu_{\textrm{rep}})|^{2}>}}
\label{eeq:deg_relative_coherence}
\end{equation}

Fig.\,\ref{fig:SWIFTS}\textbf{b} shows the SWIFTS correlation spectrum (corresponding to the numerator of Eq.\,\ref{eeq:deg_relative_coherence}) together with the spectrum product (corresponding to the denominator of Eq.\,\ref{eeq:deg_relative_coherence}). Similar to the intermode beat spectroscopy, these results analysed with SWIFTS also shows that most of the spectral power is in the frequency comb. In addition, one can also get access to the degree of relative coherence $g_{\pm}(\omega,\omega\pm\nu_{\textrm{rep}})$, which is represented in Fig.\,\ref{fig:SWIFTS}\textbf{c} for another example. This measurement shows that the degree of relative coherence is close to unity at the two regions of the gain spectrum where the output of the laser is maximum. However, this degree of relative coherence decreases to zero when one get close to the edge of each spectral region.

Finally, Table \ref{tab:table1} shows a comparison between these two methods. While both methods give access to the relative coherence of the comb, intermode beat spectroscopy is sensitive to information at different frequencies than the comb repetition frequency and is easier to implement (no need for any phase lock loop). Complementary to intermode beat spectroscopy, SWIFTS can give access to the degree of relative coherence as well as to the relative phases between pairs of modes, which can allow to retrieve the comb time-domain profile~\cite{burghoff2015evaluating}. 
Another technique has been proposed to assess the degree of coherence of THz QCL combs. It is  based on self-mixing of the laser output which is fed back into the laser cavity after being filtered by a FTIR \cite{Wienold:2014fb}.

\begin{table*}[!htb]
	\begin{center}
	\caption{
		Comparison between intermode beat spectroscopy and SWIFTS, characterization methods developed for assessing the relative coherence of QCL combs. 
		($\tau:\textrm{interferometer delay}$, $\nu_{0}$: $\textrm{reference frequency}$)} 
		\label{tab:table1}%
	\begin{ruledtabular}
		\begin{tabular}{|c|c|c|}
			 & Intermode beat spectroscopy & SWIFTS\\
			\hline
			\multirow{2}{*}{Measured quantity} & \multirow{2}{*}{$\sqrt{S_{I}^2(\tau,\nu_{0})+S_{Q}^2(\tau,\nu_{0})}$} & $S_{I}(\tau,\omega_{0}) = <(E(t)+E(t+\tau))^{2}\textrm{cos}(\nu_{0}t)>$ \\
			& & $S_{Q}(\tau,\nu_{0}) = <(E(t)+E(t+\tau))^{2}\textrm{sin}(\nu_{0}t)>$ \\ 
			\hline
			Range of $\nu_{0}$ & Spectrum analyzer span (centered at $\nu_{\textrm{rep}}$)  & Only a single frequency (usually $\nu_{\textrm{rep}}$)  \\
			\hline
			Frequency resolution & Spectrum analyzer resolution bandwidth & Lock-in bandwidth  \\
			\hline
			Sensitivity to incoherent radiation & Yes & No \\
			\hline
			Phase retrieval possible & No & Yes (cummulative sum) \\
		\end{tabular}
	\end{ruledtabular}
  \end{center}
\end{table*}

\begin{figure}[!htb]
		\includegraphics[width=0.45\textwidth]{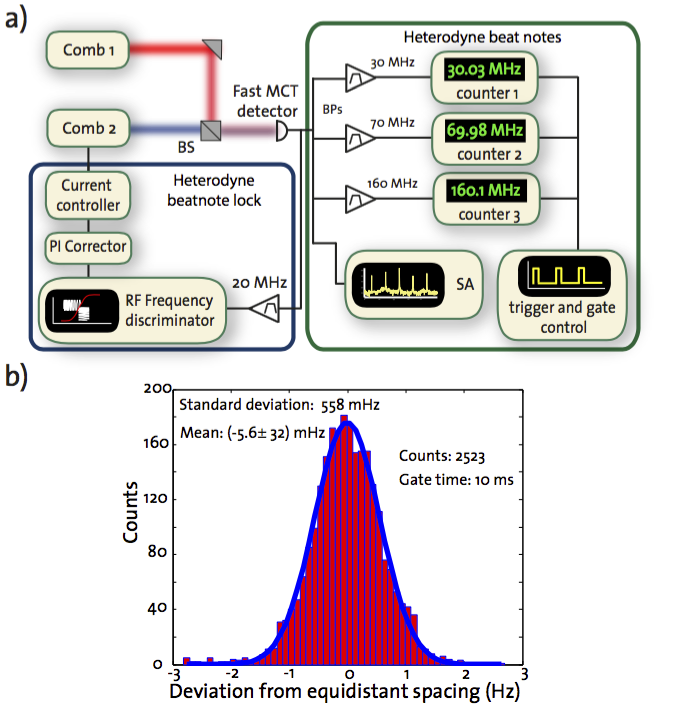}
		\caption{
			Measurement of the deviation from equidistant mode spacing.
			\textbf{a} Schematic representation of the setup used for measuring the deviation from equidistant mode spacing $\epsilon$ based on a multi-heterodyne measurement. SA: spectrum analyzer, BPs: band-pass filters, BS: beam splitter, PI: proportional integral, MCT: Mercury cadmium telluride. 
			\textbf{b} Distribution of the deviation from equidistance mode spacing $\epsilon$. Total measurement time = 25.23\,s. Gate time = 10\,ms. The distribution shows a gaussian distribution with an average value of  $(-5.6 \pm 32)$\,mHz and a standard deviation of 558\,mHz. Results published in~\cite{Villares:2014gl}.
		}
		\label{fig:Comb_equidistance}
	\end{figure}

\paragraph*{Comb line equidistance using a dual-comb technique} 
 A direct verification of the uniformity of the comb line spacing can be done by using a multi-heterodyne setup (also called dual-comb setup), as shown schematically in Fig.\,\ref{fig:Comb_equidistance}\textbf{a}~\cite{Villares:2014gl}. By employing two frequency combs with slightly different repetition frequencies, this setup effectively maps down the frequency comb from the optical domain to the RF frequencies, where established counting techniques can be employed. 
The uniformity of the mode spacing can be quantified by evaluating the deviation from equidistant mode spacing $\epsilon$, defined as~\cite{DelHaye:2014de} :
\begin{equation}
	\epsilon = \frac{f_M-f_0}{M} - \frac{f_N-f_0}{N}.
	\label{eeq:mode_spacing}
\end{equation}
where $f_0$, $f_N$ and $f_M$ are the frequencies of three multi-heterodyne beatnotes. 
After detecting the multi-heterodyne beat signal, three electrical bandpass filters are used to isolate the beatnotes at frequencies $f_0$, $f_N$ and $f_M$. After amplification, three frequency counters are used to measure the frequency oscillations of the beatnotes. 

In this experiment, an active stabilization is implemented (by acting on the current of one device) in order to correct the slow frequency drifts of the multi-heterodyne beat spectrum, thus preventing each beatnote from drifting out of the bandwidth of the electrical filters. A precise control of both trigger and gate time of all the counters was implemented. It is particularly important as any drift of $f_{\textrm{ceo}}$ will be seen as a non-uniformity of the comb line spacing if the counters are not perfectly synchronized. Even though the comb is not fully stabilized and $f_{\textrm{rep}}$ drifts, $\epsilon$ should be zero at any point in time for a perfectly synchronized setup and a perfectly spaced comb. This method offers a direct way to measure the uniform comb spacing using counting techniques even if the frequency comb is not fully stabilized. 

Fig.\,\ref{fig:Comb_equidistance}\textbf{b} shows the distribution of $\epsilon$ over the entire measurement (gate time = 10\,ms, 2523 counts), showing a mean value of $\overline{\epsilon} = (-5.6 \pm 32)$\,mHz. Normalized to the optical carrier frequency (42.8\,THz or 7\,$\mu m$), this gives an accuracy of the equidistance of $7.5$x$10^{-16}$. This measurement definitely proves the equidistance of the mode spacing of a QCL frequency comb, confirming the high degree of relative coherence of QCL combs~\cite{Villares:2014gl}.  

\paragraph*{Schawlow-Townes linewidth of QCL combs}	
As QCL combs are attractive for applications such as high-resolution molecular spectroscopy and optical metrology in the mid-infrared or THz, it is important to assess the frequency noise characteristics of these type of sources. 

\begin{figure}[!htb]
		\includegraphics[width=0.45\textwidth]{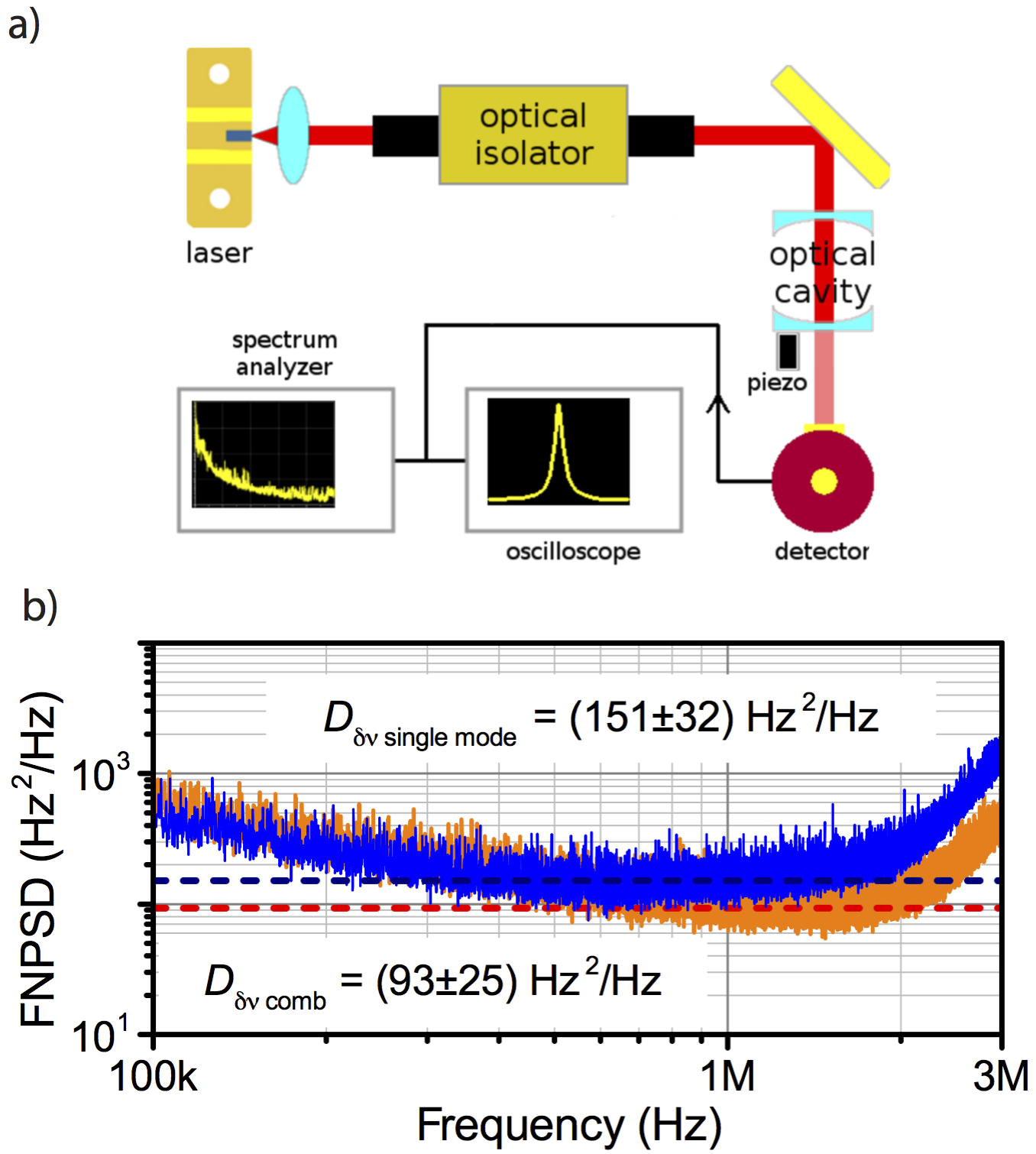}
		\caption{Schawlow-Townes linewidth of QCL combs.
			\textbf{a} Experimental setup used to measure the FNPSD of the QCL comb. The main optical components include the QCL laser, the optical isolator, the high-finesse optical cavity and the high-sensitivity MCT detector. The signal is processed by a high-sampling-rate oscilloscope.
			\textbf{b} Zoom of the flattening portion of the FNPSD of the QCL comb around 1\,MHz, corresponding to the Schawlow-Townes limit. The spectra are compensated for the FA converter cutoff. The spectra are related to the two operating conditions of the laser: single-mode with $P=15$\,mW (blue) and comb regime (with all the modes in resonance with the cavity) with $P=25$\,mW (orange). Results published in~\cite{Cappelli:2015kf}.
		}
		\label{fig:FNPSD_QCL_combs}
	\end{figure}

With this in mind, the investigation of the frequency noise power spectral density (FNPSD) of mid-infrared QCL combs was recently performed~\cite{Cappelli:2015kf}. A high-finesse optical cavity is used to resolve the laser spectrum and to detect the frequency fluctuations of the laser, acting as frequency-to-amplitude (FA) converter, as shown in Fig.\,\ref{fig:FNPSD_QCL_combs}\textbf{a}. The distance between the two mirrors is chosen in order to set the free spectral range ($FSR$) of the cavity close to comb repetition frequency $f_{\textrm{rep}}$. 
To utilize the cavity as a FA converter, a piezoelectric actuator controlling the cavity length is used and the temperature of the laser is precisely controlled in order to set the $FSR$ to match exactly the round trip frequency of the comb, while simultaneously let the comb offset frequency $f_{\textrm{ceo}}$ be equal to that of the optical cavity. In this way, the comb modes and the optical cavity resonances are perfectly matched. Successfully achieving this alignment requires an independent control of the $f_{\textrm{ceo}}$ and of the $f_{\textrm{rep}}$ of the QCL comb. 
As a consequence, in these conditions and \emph{only} in these conditions of temperature and driving current of the laser, all the comb modes are transmitted by the cavity. The cavity can thus be used as a \emph{multimode} frequency-to-amplitude converter to collect the frequency fluctuations of all the modes at the same time~\cite{Galli:2013d}. The laser emits a power of 25\,mW when the comb modes are exactly matched to the cavity resonances. A spectrum retrieved with the laser in single-mode operation ($P=15$\,mW) can also be acquired.

The FNPSD measured on the single-mode and comb regimes can therefore be measured by employing this multimode FA. Fig.\,\ref{fig:FNPSD_QCL_combs}\textbf{b} shows a magnified view of the FNPSDs in both regimes (shown from 100\,kHz to 3\,MHz). Around 1\,MHz, a flattening characteristic of a white frequency noise is observed, corresponding to the intrinsic quantum noise level $D_{\delta \nu}$ due to the spontaneous emission, the so-called \emph{Schawlow-Townes} limit~\cite{Schawlow:1958}. One can also compare these levels of $D_{\delta \nu}$ to those expected for single-mode emission with the same characteristics, given by the Schawlow-Townes limit~\cite{Henry:1982}
\begin{equation}
\delta \nu = \frac{h \nu}{P} \frac{\alpha_{\textrm{tot}} c^2}{4 \pi n_{\textrm{g}}^2} \alpha_{\textrm{m}} n_{\textrm{sp}} (1+\alpha_{\textrm{e}}^2)
\label{eq:SchToDn}
\end{equation}
Taking $\nu=42.2$\,THz as central frequency, 
$\alpha_{\textrm{m}}=2.2~\textrm{cm}^{-1}$ as mirror losses, $\alpha_{\textrm{tot}}=7.2~\textrm{cm}^{-1}$ as total cavity losses, $n_{\textrm{g}}=3.4$, $n_{\textrm{sp}}=2$ as spontaneous emission factor and $\langle \alpha_{\textrm{e}}^2 \rangle = 0.0023$ as squared \emph{Henry linewidth enhancement factor} averaged over the laser spectrum, we can compute the Schawlow-Townes limit relative to the single-mode emission ($P=15$~mW) and to the comb emission ($P=25$\,mW). The two values are $\delta \nu = 383$\,Hz and $\delta \nu = 230$\,Hz respectively. These values are consistent with those obtained from the spectra $\delta \nu = \pi D_{\delta \nu}$, which are $(474 \pm 100)$\,Hz for the single-mode emission and $(292 \pm 79)$\,Hz for the comb emission (see Fig.\,\ref{fig:FNPSD_QCL_combs}\textbf{b}).

The measurement of the FNPSD in comb regime shows that the quantum fluctuations of the different modes are correlated. In fact, we observe that the FNPSD -- in particular the portion limited by the quantum noise -- is identical when measured with one comb mode and with all comb modes simultaneously. This quantum limit, which is given by the Schawlow-Townes expression, would be at least a factor of 6 larger than the one shown in Fig.\,\ref{fig:FNPSD_QCL_combs}\textbf{b}, if we were to assume that the quantum fluctuations of each comb mode were uncorrelated. This factor is outside the uncertainty of the measurement. This experimental work demonstrate that the four-wave mixing process -- at the origin of the comb operation in QCLs -- correlates the frequency fluctuations between the modes until the quantum limit. As a consequence, instruments using the spectral multiplexing of dual-combs or multi-heterodyne spectrometers hold an inherent noise advantage compared to similar systems using arrays of single-mode lasers~\cite{Cappelli:2015kf}.

\section{Dispersion compensation}
As discussed in the previous sections, a broad gain of the QCL ensures low enough GVD to have comb operation. Nevertheless, dispersion is still large enough to prevent the comb from working over the full dynamical range of the laser, as can be seen for example in Fig.\,\ref{fig:QCL_comb_regimes} and Fig.\,\ref{fig:THz_Octave_Spanning}, and is typical in QCL comb formation~\cite{Hugi:2012ep,Burghoff:NATPHOT:2014,Villares:2014gl,burghoff2015evaluating,Cappelli:2015kf,Rosch:2014ft}. It is therefore of a major interest to compensate for this dispersion to get comb operation over the entire spectral bandwidth of the laser. In order to do such a compensation, it is crucial to have a knowledge of the exact amount of dispersion present in the laser cavity. Different methods exist for measuring the dispersion of a QCL. A straight-forward approach is to extract the mode-spacing from a high-resolution FTIR spectrum and calculate the GVD from it. This approach has proven to give an estimate of the dispersion in THz QCLs~\cite{Rosch:2014ft}. Another approach is to use THz time-domain spectroscopy to measure the phase difference between a pulse that travels once and three times through the laser cavity~\cite{Burghoff:NATPHOT:2014}. Using this phases one can calculate the GVD of the investigated laser. The theoretically predicted and measured dispersion in QCLs is positive, which requires additional negative dispersion to compensate for it.

A well-known technique in ultrafast physics to compensate for positive dispersion over broad frequency ranges and achiving octave-spanning combs are double-chirped mirrors (DCM) \cite{Matuschek:1998wb}. The working principle of a DCM is to delay different frequencies with respect to each other adding different phases for different frequencies. By carefully engeneering one can introduce a negative dispersion which exactly compensates for the intrinsic dispersion of the laser. All that is needed is a sequence of two layers with different refractive index. The thickness of one period fulfills the Bragg condition as in a Bragg mirror. The layer thickness is then chirped in a way that longer wavelengths penetrate deeper into the DCM and are therefore delayed with respect to shorter wavelengths. In addition, the duty cycle is also chirped to impedance match the DCM and avoid unnecessary oscillation in the introduced group delay \cite{Matuschek:1998wb}.

\begin{figure}[!htb]
	\includegraphics[width=0.45\textwidth]{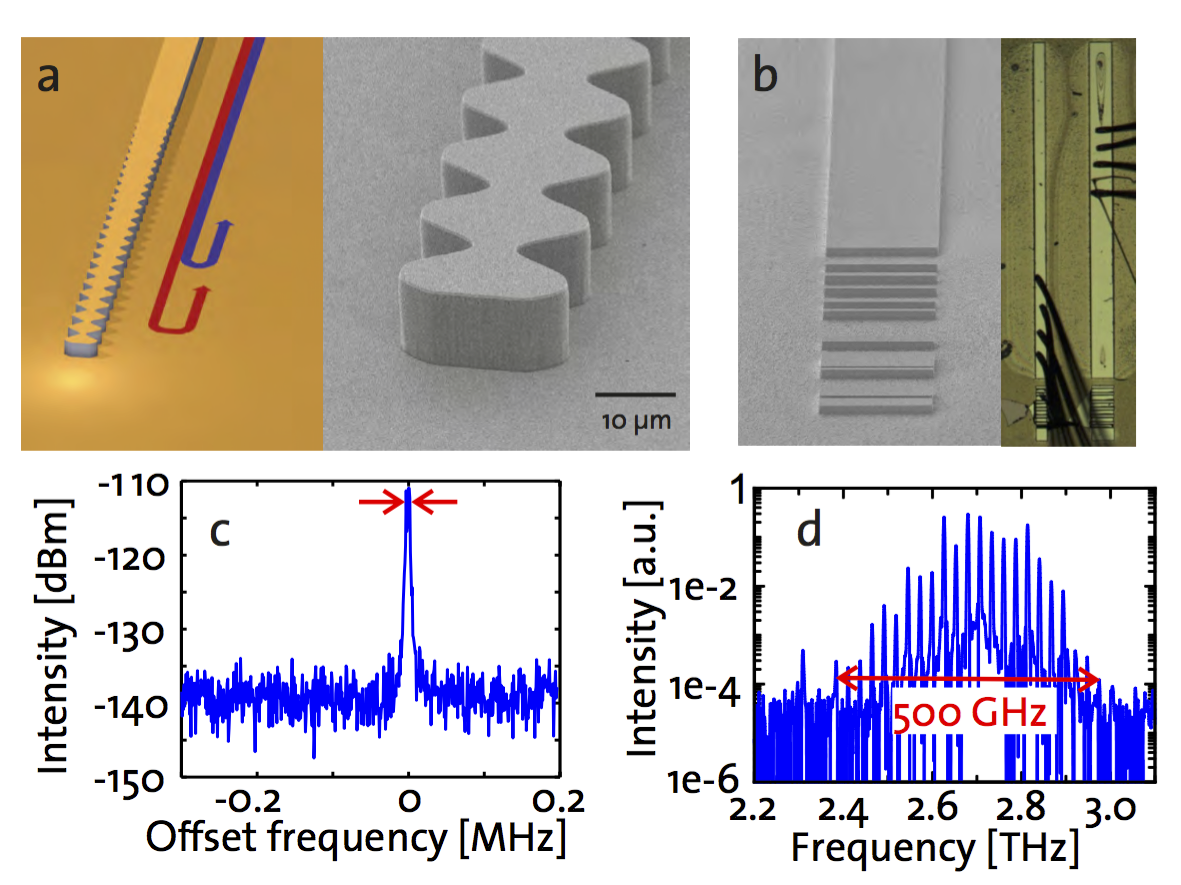}
	\caption{
		Double-chirped mirror architectures for THz QCLs: \textbf{a} Approach published by the MIT group in ref.~\cite{Burghoff:NATPHOT:2014}: The waveguide width is varied to achieve a contrast in refractive index. Both period and amplitude are chirped to introduce anomalous dispersion over a large frequency range. The longer the corrugation length the more dispersion is added. The right figure shows an SEM picture of a fabricated structure from ref. \cite{Burghoff:NATPHOT:2014} \textbf{b} Our approach, where the mirror is realized by intersecting sections of BCB and active region (MQW) to achieve a high refractive index contrast. \textbf{c} Beatnote and \textbf{d} corresponding optical spectrum of a DCM (ETHZ design) in CW operation at 15 Kelvin indicating comb operation over a spectral bandwidth of 500 GHz.
	}
	\label{fig:BCB}
\end{figure}

\paragraph*{Terahertz devices}
 Recent work introduced DCMs to compensate the dispersion in THz QCLs~\cite{Burghoff:NATPHOT:2014}. Since the dimension of DCMs is given by the Bragg condition such structures can be fabricated at THz frequencies by using standard microfabrication techniques. Instead of using two materials with different refractive index, the waveguide width has been tapered in the implementation reported in ref.~\cite{Burghoff:NATPHOT:2014}. Since the effective index of the waveguide is strongly dependent on the width, such approach is equivalent to using two media with different refractive indeces. This scheme is shown in Fig.\,\ref{fig:BCB}\textbf{a}. Starting and stoping period of the tapering are defining the frequency range covered by the DCM. The corrugation length is defining the amount of dispersion introduced by the DCM. The optical spectrum obtained with such device is shown in Fig.\,\ref{fig:SWIFTS}\textbf{b}~\cite{Burghoff:NATPHOT:2014}. 
 
A similar approach can be used to compensate for the dispersion of an octave-spanning THz QCLs. A first version has recently been realized. In order to fully exploit the technique of DCMs, sections of benzocyclobutene (BCB) and semiconductor active material have been alternated to get a higher refractive index contrast than just corrugating the waveguide as in ref.~\cite{Burghoff:NATPHOT:2014}. The top metallization is kept  continuous over the entire structure to get a good mode confinement. A series of such lasers has been realized with this technique (Fig.\,\ref{fig:BCB}\textbf{b}). As shown in Fig.\,\ref{fig:BCB}\textbf{c},\textbf{d}, a narrow beatnote is observed on such devices indicating comb operation over a bandwidth of 500\,GHz. Further optimization on the design of the DCM should allow to also compensate for a full octave in frequency. 
\paragraph*{Mid-infrared: Gires-Tournois coatings}
Another concept that can be implemented to compensate for dispersion is the use of a Gires-Tournois Interferometer~\cite{GTI_original} (GTI) directly integrated into the QCL comb. In contrast to DCMs, it can only compensate a limited frequency range but gives more flexibility on the amount of dispersion that can be introduced. This method was recently employed for the dispersion compensation of mid-infrared QCL combs, by directly evaporating the GTI on the back-facet of a QCL comb~\cite{Villares:2015ti}. By controlling the dispersion, the range where the comb operates significantly increases, effectively suppressing the high-phase noise regime usually observed in QCL combs~\cite{Hugi:2012ep,Burghoff:NATPHOT:2014,Villares:2014gl,burghoff2015evaluating,Cappelli:2015kf}. In particular, the comb regime was observed over the full dynamical range of operation of the device up to powers larger than 100 mW. 

\section{Applications: QCL-based dual-comb spectrometers}	

\begin{figure}[!htb]
 		\includegraphics[width=0.45\textwidth]{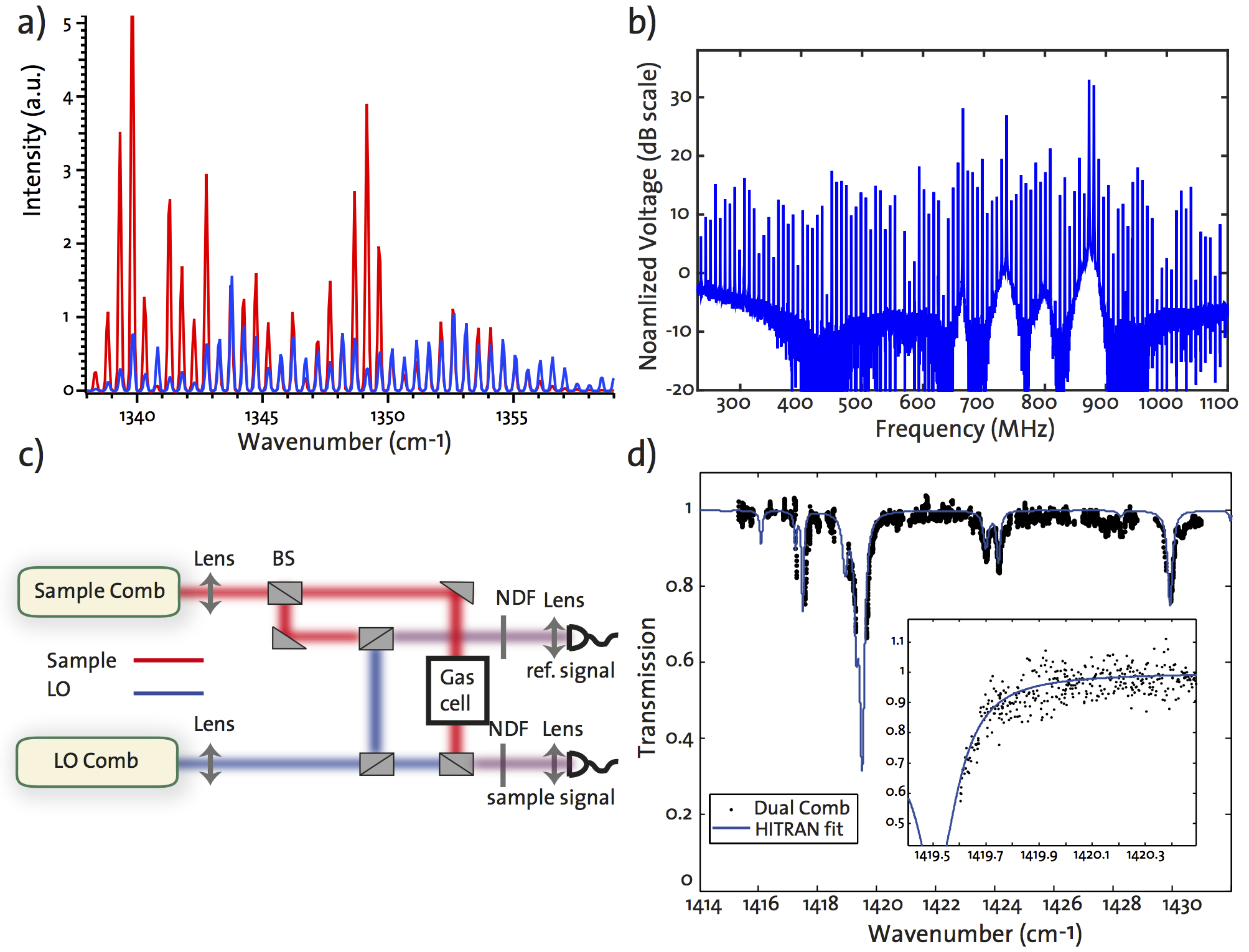}
 		\caption{ 
 			Dual-comb spectroscopy based on QCL combs:
 			\textbf{a} Typical optical spectra of two mid-infrared QCL combs, where the offset frequencies $f_{\textrm{ceo}}$ of both combs are identical. 
 			\textbf{b} Multi-heterodyne beat of two QCL combs with slightly different comb repetition frequencies measured on a fast detector. The multi-heterodyne beat signal contains information on the sample absorption.
 			\textbf{c} Schematic view of the dual-comb spectroscopy setup based on QCL frequency combs. One comb is used as a local oscillator (LO) while the other interrogates the gas cell.  BS: 50-50 Antireflection coated beam splitter, NDF: neutral density filter.
 			\textbf{d} Water vapour transmission spectra in air ($P_{\textrm{H}_{2}\textrm{O}} = 1.63$\,kPa, total pressure $P_{\textrm{tot}} = 101.3$\,kPa, $T = 20^{\circ}$C) measured with our dual-comb spectrometer (800\,MHz of spectral resolution after averaging) and HITRAN simulation. Inset: expansion of the measured transmission over $\sim$1\,cm$^{-1}$ with the full resolution (80\,MHz). Results published in~\cite{Villares:2014gl}.
 		}
 		\label{fig:Dual_comb_trans_meas}
 	\end{figure}

While frequency combs can be used for spectroscopy by combining them with dispersive elements~\cite{Diddams:2010p2046}, the most attractive use of these devices is clearly the multi-heterodyne or dual-comb configuration~\cite{Keilmann:2004p2040,Coddington:2008p99}, because it leverages on the unique coherence properties of the comb. As schematically described in Fig.\,\ref{fig:dualcomb} and as shown in in Fig.\,\ref{fig:Dual_comb_trans_meas}\textbf{c} for an implementation using QCL combs, dual-comb spectroscopy is based on the measurement of a beating created by two frequency combs with slightly different repetition frequencies ($f_{\textrm{rep},1}$ and $f_{\textrm{rep},2} = f_{\textrm{rep},1} + \Delta f$, respectively, where $\Delta f $ is the difference in the combs repetition frequencies). 

Results using this technique combined with QCL combs, published in ref.~\cite{Villares:2014gl}, are summarized in Fig.\,\ref{fig:Dual_comb_trans_meas}. The optical spectrum of two QCL combs tuned to allow multiheterodyne spectroscopy is shown in Fig.\,\ref{fig:Dual_comb_trans_meas} \textbf{a}, while the RF spectrum associated with the beating of two QCL combs is shown in Fig.\,\ref{fig:Dual_comb_trans_meas}\textbf{b}. These key results shows the principle of dual-comb spectroscopy.  A detector is used to measure the multi-heterodyne beat, corresponding to Fig.\,\ref{fig:Dual_comb_trans_meas}\textbf{b}. Each line of the first comb will beat with all lines of the second comb creating several different beatnotes in the RF domain. The difference in repetitions frequencies $\Delta f$ must be chosen such as to obtain a one-to-one mapping between the lines of the two combs and their RF counterpart.
 
 In one form of dual-comb spectroscopy~\cite{coddington2010coherent}, one comb is used as a local oscillator while the other is used to interrogate a sample, as shown in Fig.\,\ref{fig:Dual_comb_trans_meas}\textbf{c}. Due to the one-to-one mapping, each multi-heterodyne beat contains information regarding the sample absorption at the optical frequency of the comb line interrogating the sample. As the technique relies on the discrete nature of a frequency comb, the sample absorption is measured at frequencies spaced by the comb repetition frequency ($7.5$\,GHz $= 0.25$\,cm$^{-1}$ for the example of Fig.\,\ref{fig:Dual_comb_trans_meas}), thus defining the resolution of the dual-comb spectrometer. For the investigation of large organic molecules in gas phases or for the study of liquids, this resolution is usually sufficient. However, for gas spectroscopy of small molecules at low pressures, the linewidth of molecule absorption lines are usually narrower (hundreds of MHz) than QCL comb repetition frequencies, usually ranging from 5\,GHz up to 50 GHz for conventional devices. A special feature of QCL combs is that the comb teeths can be swept over an entire $f_{\textrm{rep}}$, by tuning either the laser temperature or current, therefore increasing the resolution of the QCL-based dual-comb spectrometer.  

A dual-comb spectrometer based on QCL combs is depicted in Fig.\,\ref{fig:Dual_comb_trans_meas}\textbf{c}. The difference in repetitions frequencies $\Delta f $ can be set between 5 to 40\,MHz by changing the temperature and current of both combs. A dual detection technique is implemented as it helps to remove technical noise on the detected amplitude \cite{newbury2010sensitivity}. As a proof of principle applied to gas sensing, transmission measurement of water vapour in air at atmospheric pressure in a 6\,cm-long gas cell is measured. We applied a frequency sweep to the combs in order to achieve high-resolution (80\,MHz step over a bandwidth equal to one comb repetition frequency). In order to avoid parasitic fringes coming from residual reflectivities in the beam path, the multi-heterodyne beat signal is measured with the gas cell filled with water vapour and subsequently with nitrogen, at each step of the frequency sweep. A reference measurement was thus taken at each step of the sweep and used to deduce the absolute value of the transmission.

Fig.\,\ref{fig:Dual_comb_trans_meas}\textbf{d} shows the transmission of water vapour in air (partial pressure of water vapour in air $P_{\textrm{H}_{2}\textrm{O}} = 1.63$\,kPa, total pressure $P_{\textrm{tot}} = 101.3$\,kPa, $T = 20^{\circ}$C) measured with the dual-comb setup as well as a transmission simulation using HITRAN database \cite{rothman2009hitran}. The wavenumber scale calibration was done by using the HITRAN simulation and by applying a rigid shift to the measured transmission spectrum in order to fit the HITRAN simulation spectrum. Over the entire duration of the measurement (few hours), the slow drifts reduce the quality of the interleaved transmission spectrum and a moving average filter was used for smoothing the interleaved transmission spectrum, reducing the effective resolution to 800\,MHz.

A clear agreement between the two measurements is observed over the entire measurement bandwidth (16\,cm$^{-1}$) and the absolute value of the transmission could also be retrieved. Due to the important attenuation of the comb lines situated in the vicinity of the water absorption lines, their attenuated amplitude could lie below the noise floor of the detection setup. In such a case, the algorithm calculating the absorption will not be able to retrieve the entire shape of the absorption line. This can be observed on the shape of the water absorption line situated at 1419.5\,cm$^{-1}$. Finally, as the multi-heterodyne spectrum presents beatnotes with difference in amplitudes of more than 40\,dB, the very low intensity beatnotes will not be detected by the algorithm calculating the absorption. This results in some holes on the spectrum and can also be observed on Fig.\,\ref{fig:Dual_comb_trans_meas}\textbf{c}, for instance close to the water absorption line situated at 1416.3\,cm$^{-1}$. 

%
\begin{figure}[!htb]
 		\includegraphics[width=0.5\textwidth]{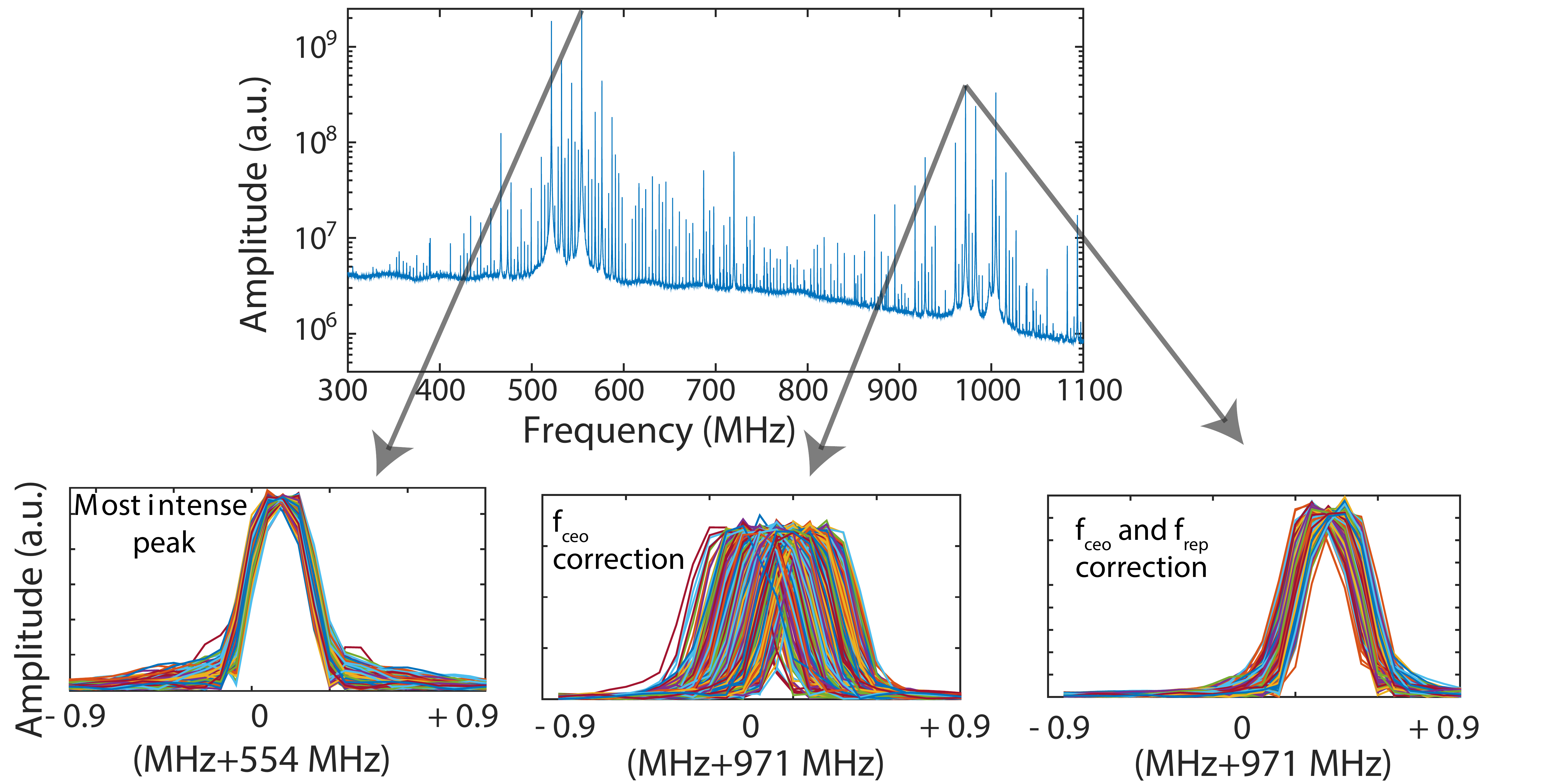}
 		\caption{ 
 			Digital adaptive sampling technique used to correct $f_{\textrm{ceo}}$ and $f_{\textrm{rep}}$ fluctuations. The improved algorithm corrects for both $f_{\textrm{ceo}}$ and $f_{\textrm{rep}}$.
			}
 		\label{fig:algorithm}
\end{figure}

In general, the signal over noise characteristics of the absorption measurements in free-running dual-comb spectroscopy reflects a trade-off between the spectral bandwidth, number of comb lines and amplitude accuracy. Using conventional Fabry-P\'erot QCL devices, dual-comb spectroscopy was demonstrated successfully on a relatively limited spectral bandwidth~\cite{Wang:2014hca}. In fact, the observation of the Schawlow-Townes limit on the linewidth of QCL combs proves that the comb formation process does not add additional mode partition noise; what must be done is to remove the fluctuations in the $f_{\textrm{ceo}}$ and $f_{\textrm{rep}}$ in both combs. In conventional dual-comb systems, such noise has been efficiently removed by "adaptative sampling" techniques~\cite{Ideguchi:2014jb}. 

Such adaptive sampling techniques can be fitted to QCL combs.
The results presented in Fig.\,\ref{fig:Dual_comb_trans_meas}\textbf{d} were achieved by correcting fluctuations in $f_{\textrm{ceo}}$ but neglected the fluctuations of $f_{\textrm{rep}}$. The spectra were aligned spectrally along the main peak; as a of fluctuations of the $f_{\textrm{rep}}$, the wings of the spectra suffered from a conversion of that frequency noise into an additional amplitude noise. This effect is shown in Fig.~\ref{fig:algorithm}: while the fluctuations of the multi-heterodyne beatnote at around the most intense peak ($\sim$ 554\,MHz) are corrected over the whole acquisition time, the peak at ($\sim$ 971\,MHz) is fluctuating by more than its linewidth. When $f_{\textrm{rep}}$ is simultaneously retrieved and the signal resampled, the frequency fluctuations are significantly reduced. 

\begin{figure}[!htb]
 		\includegraphics[width=0.45\textwidth]{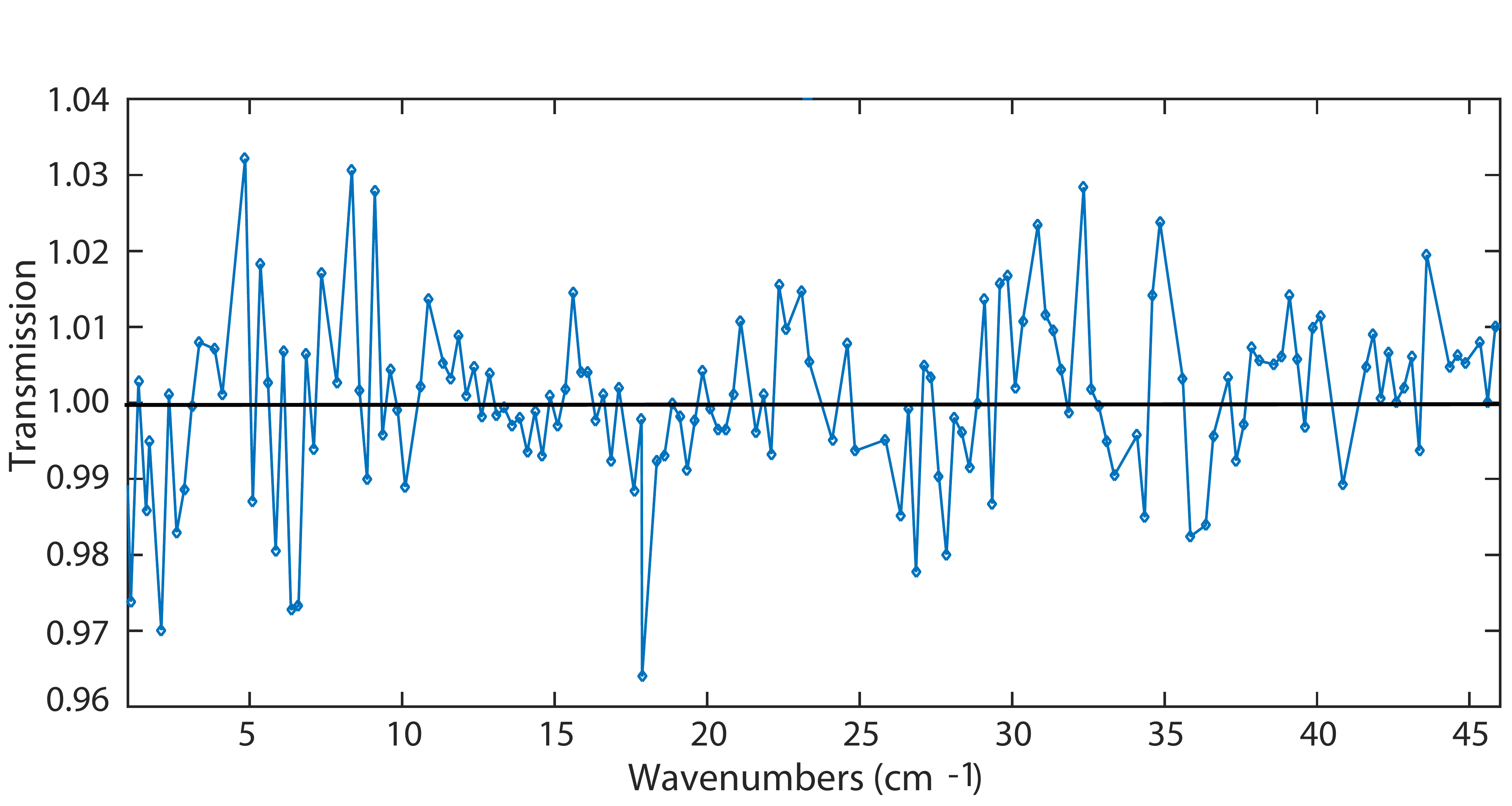}
 		\caption{ 
 			Baseline measurement of dual-comb spectrometer with both $f_{\textrm{ceo}}$ and $f_{\textrm{rep}}$ corrections. Data acquisition time is 25\,ms. 	}
 		\label{fig:baseline}
\end{figure}

The effect of this correction of both $f_{\textrm{ceo}}$ and $f_{\textrm{rep}}$ is shown in Fig.~\ref{fig:baseline} where two subsequent acquisition are ratioed, as an indication of the instrument baseline. The measured SNR of each measurement point varies according to the relative intensity of the peak. In an acquisition time of 25\,ms, SNR of <$1\times10^{-3}$ of high power peaks are reached whereas low power peaks exhibit a SNR of <$4\times10^{-2}$. Compared to the situation where only $f_{\textrm{ceo}}$ is corrected, the useful bandwidth of the dual-comb spectrometer is increased from 16 to 45\,cm$^{-1}$.
%
%

Similar, preliminary results were recently presented, demonstrating the use of correction techniques to THz combs~\cite{TowardsTHzdualcom:2015wy}. As such, although self-referencing is a desirable property of such combs ultimately, it is probably not a requirement for the development of  high performance systems. 
%

\section{Conclusion and outlook}
 As compared to mode-locked monolithic semiconductor lasers, QCL combs benefits from features specific to the physics of its intersubband gain medium: the capability to achieve very wide gain bandwidths, the low material GVD as well as the FM-like characteristics of the emission that allows large average powers without high peak powers incompatible with tightly confined waveguides. The possibility to build broadband spectrometers with chip-sized, electrically pumped optical frequency comb sources has a great potential for applications in sensing and spectroscopy. QCL combs have, in addition, some additional potential features that have not yet been exploited: because of the ultrafast transport in the active region, QCLs are also RF detectors that are in principle capable of detecting a heterodyne beatnote. The recent progress in broadband gain active regions in the mid-infrared and THz show that an octave-spanning QCL comb is feasible. In addition, the active region can also be engineered to provide a $\chi^{(2)}$ susceptibility, and therefore internally generate its own second harmonic. This capability of the QCL comb to integrate gain, non-linearity as well as detection is unique and offers great potential for device and system integration.  
 
\section{Acknowledgement}
The authors want to acknowledge the support from the Swiss National Science Fundation, the ETH pioneer grant, the FP7 project TERACOMB as well as from the DARPA program SCOUT. 
 
\bibliographystyle{apsrev}
\bibliography{bibtex-library,MyLibrary,MyLibraryGS,MyLibraryGV}
\end{document}